\documentclass[conference]{IEEEtran}
\IEEEoverridecommandlockouts
\usepackage{cite}
\usepackage{amsmath,amssymb,amsfonts}
\usepackage{algorithmic}
\usepackage{graphicx}
\usepackage{subcaption}
\usepackage{textcomp}
\usepackage{xcolor}
\usepackage{comment}
\usepackage{enumitem}
\usepackage[draft]{hyperref}
\usepackage{xurl}

\usepackage[top=0.765in,bottom=1.04in,left=0.625in,right=0.625in]{geometry}

\def\BibTeX{{\rm B\kern-.05em{\sc i\kern-.025em b}\kern-.08em
    T\kern-.1667em\lower.7ex\hbox{E}\kern-.125emX}}
\begin{document}

\title{Implementation and Calibration of 3GPP-Compliant ISAC Channel Simulator}
\author{Chien-Han Wu, Ming-Chun Lee, and Ta-Sung Lee\\
Institute of Communications Engineering, National Yang Ming Chiao Tung University, Hsinchu, Taiwan\\
E-mail: \{hector313513055.ee13, mingchunlee, tslee\}@nycu.edu.tw}

\maketitle

\begin{abstract}

Integrated sensing and communication (ISAC) has emerged as a key technology for 6G systems. To support the development of ISAC systems, accurate channel modeling and simulation for performance evaluation is essential. Recently, 3GPP introduced a standardized ISAC channel model and its associated calibration procedure for this purpose. However, due to the complexity of the modeling methodology and the lack of fully explicit implementation details in the 3GPP reports, different implementations may lead to inconsistent or unsynchronized simulation results. To address this issue, in this work, we implement the 3GPP ISAC channel model simulator specified in TR 38.901 and conduct a comprehensive calibration analysis. We compare the simulation results with the reference results reported by companies in 3GPP and discuss several key implementation details to provide insights into the implementation and calibration of the simulator. To facilitate reproducibility and further research, the developed simulator, together with the relevant datasets and calibration results, has been released as an open-source project on GitHub.


\end{abstract}


\section{Introduction}

As wireless communication systems evolve toward 6G, emerging applications require not only data transmission but also accurate environmental awareness. Accordingly, integrated sensing and communication (ISAC) has emerged as a key enabling technology for these applications \cite{hong2026integrated,wei2024integrated}. 
From a signal-processing perspective, sensing relies on analyzing echo signals reflected from targets and the surrounding environment. Therefore, sensing approach design and performance evaluation are highly dependent on accurate channel modeling \cite{wei2024integrated}. Compared to conventional communication channels, ISAC channels require more detailed modeling of physical propagation effects, including geometric distances, target motion, delay and Doppler characteristics, as well as radar cross section (RCS) properties with large-scale, small-scale, and angular-dependent components \cite{zhang2025research,liu2025coupling,gomez2026geometry,yang2025general,zhang2025unified}. Hence, the ISAC channel model cannot be simply treated as a direct extension of original communication channel models.

To address the need for an accurate ISAC channel model and simulation methodology, the 3GPP RAN1 working group has recently developed a standardized ISAC channel model and its simulation procedure. Since early 2024, the model has evolved to include deployment scenarios, channel frameworks, physical object modeling, target and background channels, as well as additional sensing-specific components. The corresponding calibration methodology and reference results were then released in 2025, providing a baseline for validating the 3GPP-compliant ISAC channel simulations, where a detailed overview of the relevant evolution can be found in \cite{liu2025comprehensive,eren2025integrated}.

However, implementing the ISAC channel model and its corresponding simulator according to the 3GPP specification is a non-trivial task \cite{hong2026integrated,liu2025comprehensive}. This is because the existing ISAC modeling methodology is highly complicated, involving numerous detailed implementation procedures and a large number of parameters. As a consequence, different implementations may easily lead to highly inconsistent simulation outcomes, despite all claiming compliance with the same 3GPP specification. Hence, proper calibration plays a critical role in ensuring that the implemented channel model is consistent with the reference results provided by 3GPP~\cite{tang2026tutorial}. Unfortunately, while 3GPP TR 38.901 provides a detailed calibration setup and procedure, the explanations are limited and may easily lead to misunderstandings. Moreover, practical implementation aspects and the interpretation of calibration assumptions are commonly unclear for people who are not deeply involved in 3GPP activities. On the other hand, although some existing papers have discussed the fundamental insights and rationale behind the 3GPP ISAC channel modeling methodology, e.g., \cite{liu2025comprehensive,tang2026tutorial,eren2025integrated}, to the best of our knowledge, the detailed calibration procedure and implementation aspects of the 3GPP ISAC channel model have not been systematically discussed, which hinders the straightforward development of a 3GPP-compliant channel simulator. In addition, certain aspects of ISAC channel modeling are not specified in a sufficiently implementation-explicit manner by them. These ambiguities become particularly critical during the calibration process, where even minor differences in modeling assumptions can result in noticeable deviations from the reference results. Observing these, this paper aims to bridge the gaps by providing a detailed discussion on the implementation and calibration of the 3GPP ISAC channel model simulator. The main contributions of this paper are as follows:
\begin{itemize}
    \item We develop a 3GPP-compliant ISAC channel simulator based on TR 38.901 \cite{3gpp38901v1910} and provide detailed implementation discussions for both target-channel and background-channel generation.
    \item We conduct a comprehensive calibration study by comparing the developed simulator with the reference results reported by companies in 3GPP. The calibration results validate that the proposed simulator reproduces the intended statistical behavior of the 3GPP ISAC channel model across representative scenarios and metrics.
    \item We identify and discuss several implementation-dependent details that materially affect calibration outcomes, including sensing-target selection, monostatic coefficient construction, ZOA-based filtering, indoor target distribution, and calibration pair selection. These discussions provide practical guidance for developing and verifying interoperable 3GPP-compliant ISAC simulators.
    \item To facilitate reproducibility and further ISAC research, the developed simulator, together with the relevant datasets and calibration results, has been released as an open-source project on GitHub.\footnote{GitHub repository: \url{https://github.com/Putirf/Implementation-and-Calibration-of-3GPP-Compliant-ISAC-Channel-Model-Simulator}. The simulator, relevant datasets, and calibration results are publicly available online.}
\end{itemize}

\section{ISAC Channel Modeling Framework}

In this section, we briefly discuss the 3GPP ISAC channel modeling framework, and the explanations of the framework can be found in \cite{liu2025comprehensive,huang2023flexible}. The 3GPP ISAC channel model extends the conventional 3GPP geometry-based stochastic communication channel model by incorporating sensing-specific propagation mechanisms. Unlike a communication channel, the ISAC target channel involves a cascaded path from the sensing transmitter to a target and then to the sensing receiver. Therefore, additional procedures such as target reflection modeling, RCS generation, and ray coupling are required for ISAC channel construction. In particular, the 3GPP ISAC channel model consists of two components, namely the target channel and the background channel, expressed as:
\[
H^{\mathrm{ISAC}}(\tau,t)
=
\sum_{k} H^{(k)}(\tau,t)
+
H^{\mathrm{bk}}(\tau,t),
\]
where $H^{(k)}(\tau,t)$ is the reflected link in which the signal propagates from the transmitter to the $k$-th sensing target, and then to the receiver, and $\sum_{k} H^{(k)}(\tau,t)$ is the overall target channel that represents the superposition of propagation paths associated with multiple sensing targets (ST); the background channel $H^{\mathrm{bk}}(\tau,t)$ represents the direct propagation and environmental multi-path components between the transmitter and the receiver. In a bistatic sensing configuration, the background channel is equivalent to the conventional communication channel between the transmitter and the receiver. In contrast, in a monostatic sensing configuration, the transmitter and receiver are co-located within the same device. Therefore, no communication link between the transmitter and receiver is explicitly modeled, unlike the bistatic case. Hence, to construct the background channel under this configuration, additional reference points are introduced in the 3GPP specification to generate equivalent propagation paths.

\subsection{Target Channel Modeling}

\begin{figure}[t]
\centering
\vspace{0.08in}
\includegraphics[width=0.44\textwidth]{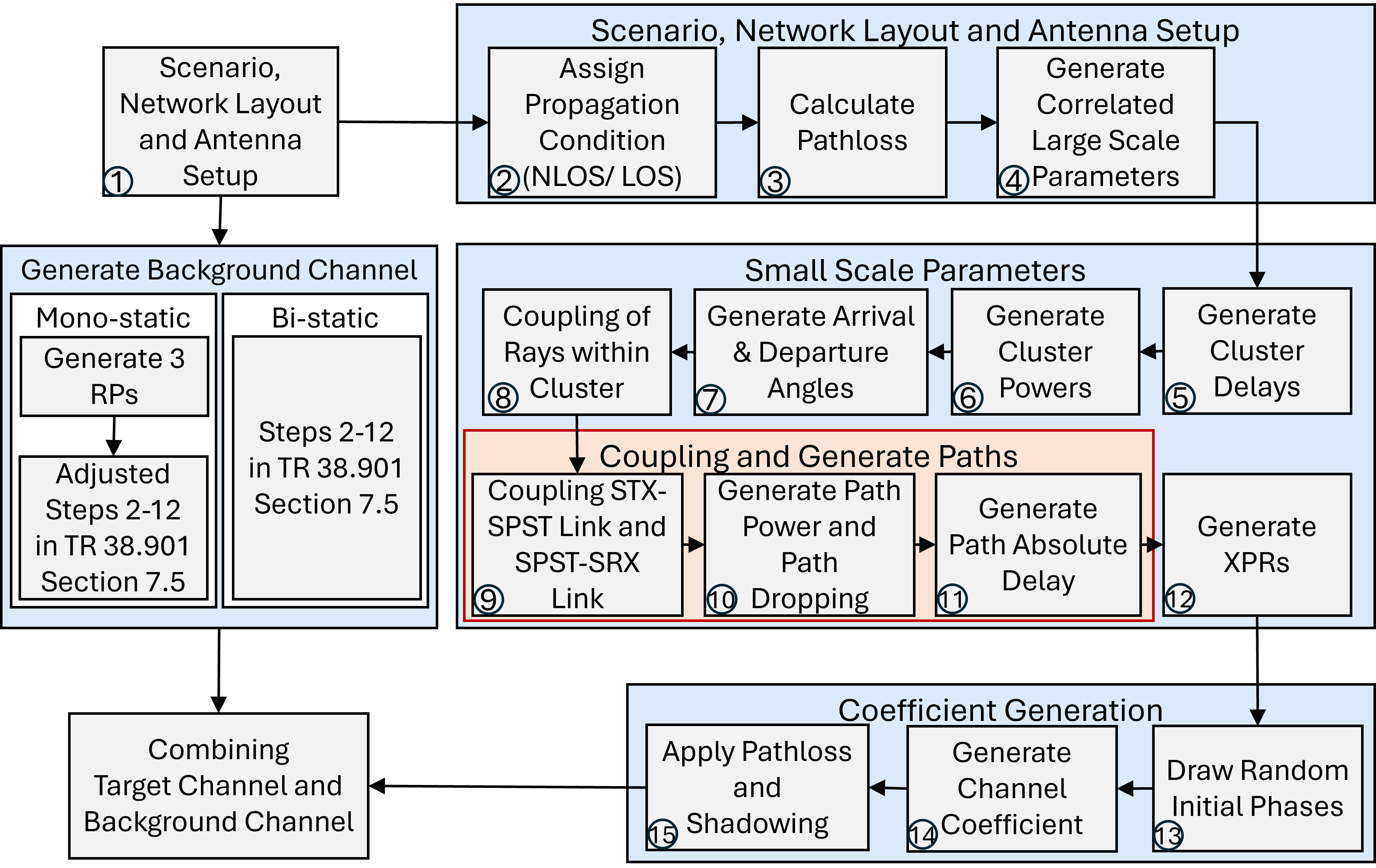}
\caption{ISAC channel coefficient generation procedure.}
\vspace{-15pt}
\label{fig:Channel coefficient generation procedure}
\end{figure}

The target channel modeling procedure is illustrated in Fig.~\ref{fig:Channel coefficient generation procedure}, where most of the steps follow the conventional communication coefficient generation procedure \cite{huang2023flexible}, with the exception that steps 9-11, highlighted in the Coupling and Generate Paths blocks, are newly introduced for the ISAC channel model. Specifically, steps 9-11 of the target channel model couple the forward and backward propagation links associated with the same scattering point of the sensing target (SPST), and then generate the corresponding path power and delay, thereby capturing the physical characteristics of the sensing process. In step 9, the line-of-sight (LOS) and non-LOS (NLOS) rays of the sensing transmitter (STX)–SPST and SPST-sensing receiver (SRX) links associated with the same SPST are coupled to form complete propagation paths. A direct path is formed when both links consist of LOS rays, whereas indirect paths arise from combinations of LOS and NLOS rays, including LOS+NLOS, NLOS+LOS, and NLOS+NLOS cases. For the NLOS+NLOS case, 3GPP TR 38.901 provides two coupling options: (i) exhaustive pairing of all NLOS rays in both links, and (ii) random pairing of $\min(M_1,M_2)$ rays, where $M_1$ and $M_2$ are the numbers of rays in the two links. These options allow different levels of modeling resolution for the NLOS-NLOS paths.

In step 10, the power of each coupled path is generated based on the ray powers of the STX to SPST and SPST to SRX links, as well as the RCSs of the STs. The RCS characterizes the effective reflecting strength of the target and is modeled using small-scale RCS components (e.g., $\sigma_D$ and $\sigma_S$) whose particular values depend on the LOS/NLOS conditions of both links. These components capture the small-scale fluctuations of the target reflection, and paths with power below a predefined threshold are discarded, except for the LOS path. In step 11, the delay of each path is determined by extending the cluster delay defined in step 6. Specifically, in addition to the cluster delay, the overall delay of a path includes the propagation delay corresponding to the 3D distance and additional delay components associated with indirect NLOS paths, as defined in Section 7.6.9 of the communication channel model. Hence, Step 11 ensures that the resulting delay accurately reflects the physical propagation characteristics of the sensing paths. Overall, steps 9-11 model the reflection and scattering effects introduced by the sensing target, enabling a physically consistent representation of the sensing channel.

Finally, the channel coefficient generation in Step 14 follows a procedure similar to that used in the communication channel model. However, it additionally incorporates the cross-polarization matrix (CPM) of the sensing target, along with phase shifts induced by delay and Doppler. The CPM characterizes the target’s polarization-dependent reflection behavior, while the phase shifts capture the target’s range and velocity information.

\subsection{Background Channel Modeling}

For a bistatic sensing configuration, the background channel is equivalent to the communication channel between the transmitter and the receiver, which is straightforward. Hence, we focus on the monostatic background channel modeling procedure. For the monostatic case, no separate receiver exists to form a link. To address this, the 3GPP channel model introduces three reference points (RPs) that act as virtual receivers. The positions of the RPs are determined by their range and height, which are jointly modeled using a combination of the Gamma distribution and a constant component. The angular directions of the RPs are then uniformly distributed around the transmitter, with azimuth angles separated by $120$ degrees.

Since the RPs are virtual receivers without physical presence, the links between the transmitter and the RPs are always treated as NLOS, and no outdoor-to-indoor (O2I) condition is considered. In addition, the RPs do not have a specific orientation. Therefore, in step 7, the angles of arrival (AoAs) are set equal to the angles of departure (AoDs) in both azimuth and elevation domains, while following the same coupling defined in step 8. Furthermore, rays within a cluster are subject to scenario-dependent filtering. Specifically, rays with zenith angles of arrival (ZOA) below 50 and 80 degrees are discarded for urban-micro (UMi) and urban-macro (UMa) scenarios, respectively. The propagation delay is determined by the 3D distance between the transmitter and the RPs, where the stochastic component of the delay is mainly governed by the Gamma-distributed range. Finally, the background channel in the monostatic case is obtained by combining the channel responses associated with all three RPs, thereby approximating the surrounding scattering environment of the transmitter.

\section{Simulation Calibration Approach}

To allow simulators implemented by different groups to be synchronized, a common calibration procedure needs to be conducted. Hence, in this section, we present the calibration methodology, including the simulation setup, key implementation details required to ensure consistency with the 3GPP specification, and the definitions of the calibration metrics. 

\subsection{Calibration Setups}

The simulation parameters for the calibration of different sensing types follow 3GPP TR 38.901, as specified in Tables 7.9.6.1-1 to 7.9.6.1-4 for large-scale calibration and Tables 7.9.6.2-1 to 7.9.6.2-4 for full calibration. Specifically, two representative frequency configurations are considered. The first corresponds to the FR1 setting with carrier frequency of $6$ GHz, bandwidth of $100$ MHz, base station (BS) transmit power of $56$ dBm, and noise figures of $5$ dB and $9$ dB for the BS and user equipment (UE), respectively. The second corresponds to the FR2 setting with carrier frequency of $30$ GHz, bandwidth of $400$ MHz, BS transmit power of $41$ dBm, and noise figures of $7$ dB and $10$ dB for BS and UE, respectively.

\begin{figure}[t]
\centering
\vspace{0.08in}
\begin{subfigure}{0.43\linewidth}
\centering
\includegraphics[width=\linewidth]{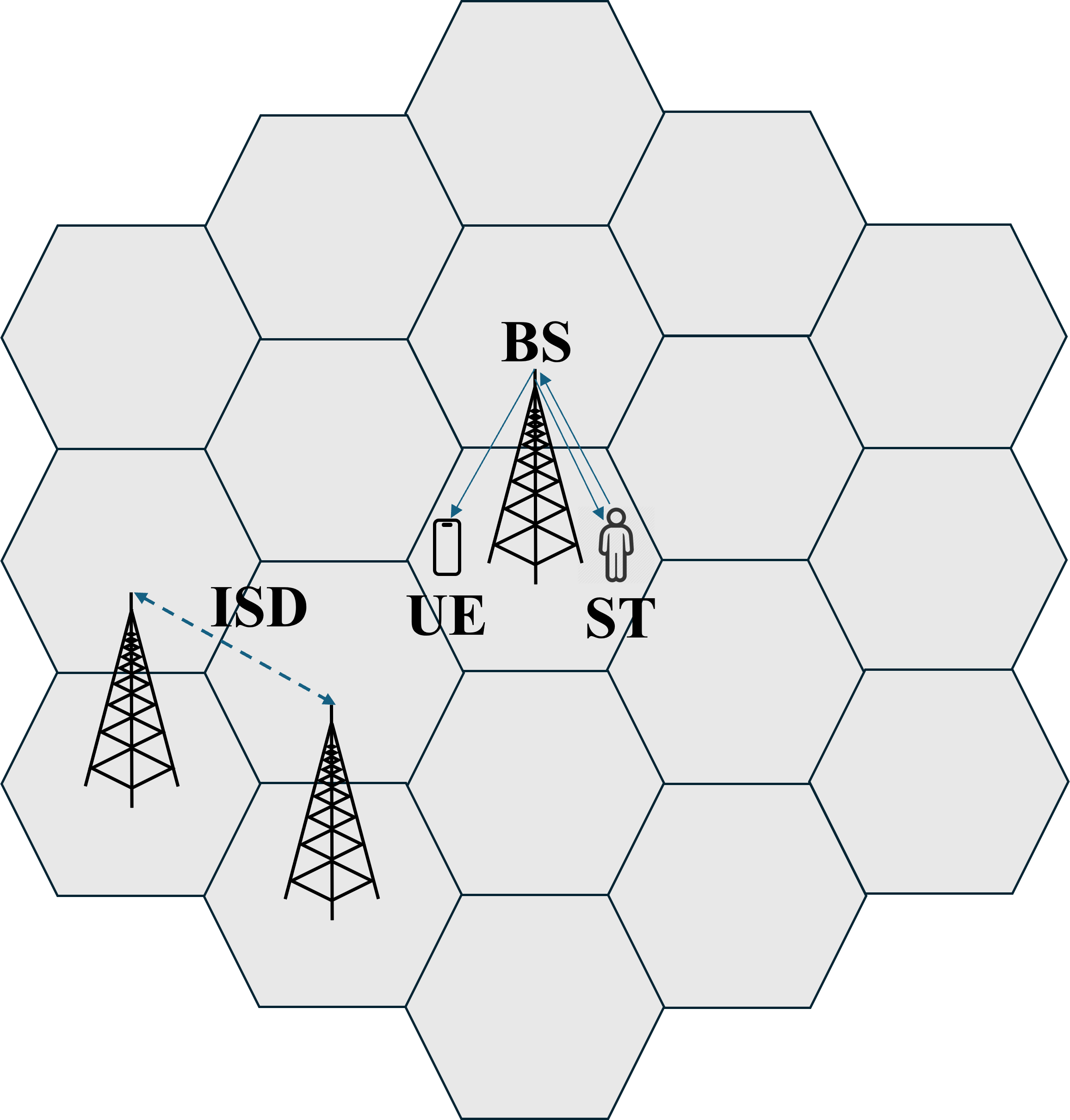}
\caption{Outdoor.}
\label{fig:The network layout of outdoor test environments}
\end{subfigure}
\hfill
\begin{subfigure}{0.45\linewidth}
\includegraphics[width=\linewidth]{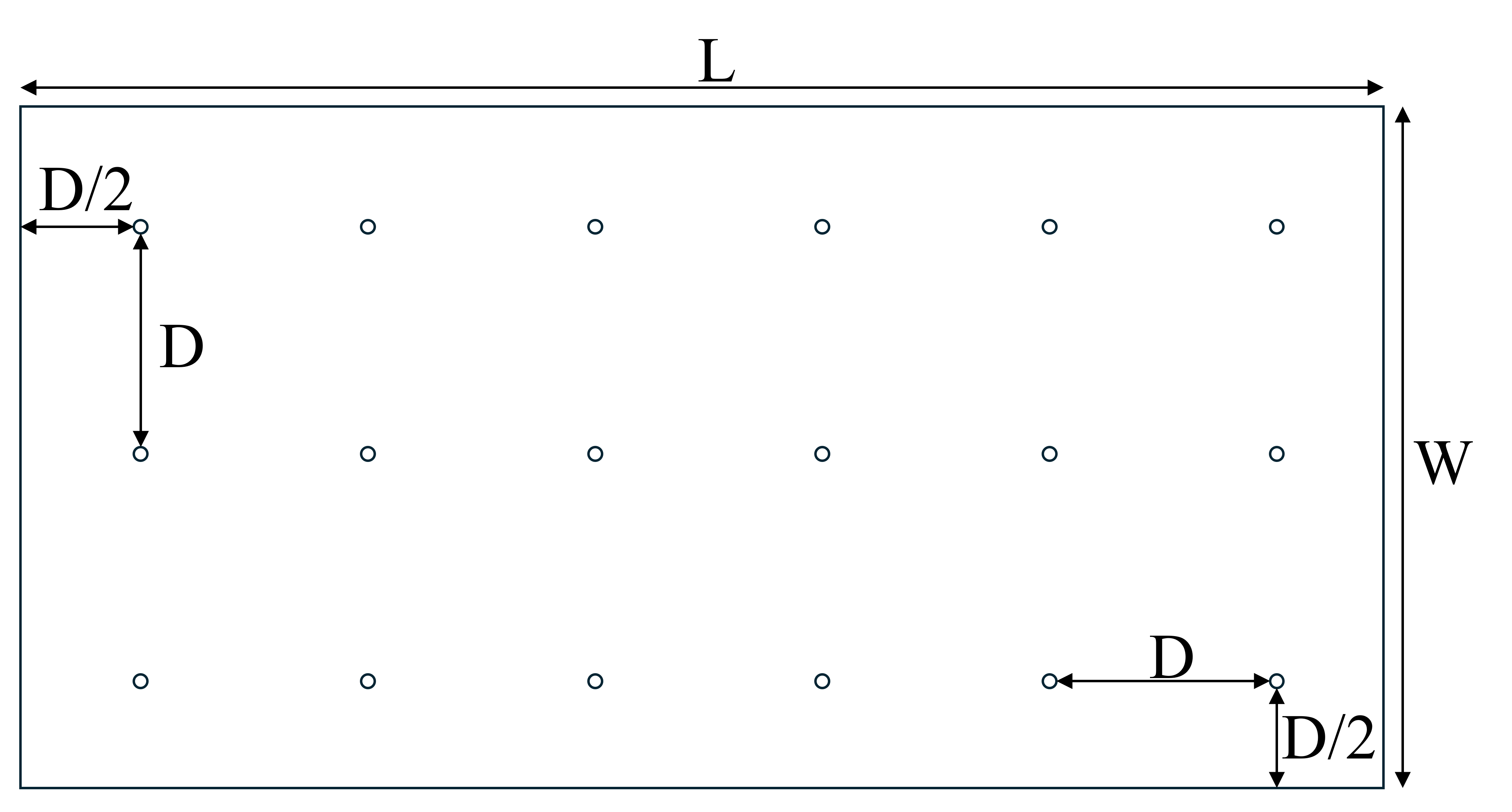}
\caption{Indoor.}
\label{fig:The network layout of indoor test environments}
\end{subfigure}


\caption{Illustration of the network layout for different scenarios.}
\vspace{-10pt}
\label{fig:compare}
\end{figure}

The calibration scenarios consist of outdoor and indoor scenarios. In the general outdoor scenario, unlike the conventional communication channel model, which assumes three sectors per cell site, the ISAC channel calibration adopts a single 360-degree sector configuration, where the BSs are located at the center of a hexagonal cell, as illustrated in Fig.~\ref{fig:The network layout of outdoor test environments}. For indoor scenarios, although 3GPP TR 38.901 defines both indoor office (InH) and indoor factory (InF) scenarios, their BS deployment follows a similar configuration, as illustrated in Fig.~\ref{fig:The network layout of indoor test environments}. In particular, the BSs are arranged on a square lattice with spacing $D$ and are positioned at a distance of
$D/2$ from the walls, where $D$ is determined by the scenario dimensions and the number of BSs. Finally, the Urban Grid scenario 
is considered for the calibration of automotive STs. This scenario includes BSs, pedestrian UEs, roadside-unit (RSU)-type UEs, and vehicle-type UEs. The detailed parameter settings can be found in 3GPP TR 37.885~\cite{3gpp37885v1530}. It should be noted that this paper focuses only on the static calibration scenarios specified in 3GPP TR 38.901; the spatial consistency and UE mobility will be incorporated in the future. 

\subsection{Calibration Details}

The calibration involves several implementation-specific considerations that are not explicitly described in the 3GPP document but are nevertheless critical for accurate calibration. The details requiring special attention are listed below.
\begin{enumerate}[label=\Roman*.]
    \item For UAV scenarios, the pathloss model specified in 3GPP TR 38.901 is defined for user terminal heights ranging from $1.5$ m to $22.5$ m. For heights exceeding $22.5$ m, the pathloss model is extended by adopting the model specified in 3GPP TR 36.777 \cite{3gpp36777v1500}.
    \item For indoor scenarios, STs are uniformly distributed over the convex hull of the BS deployment. 
    However, this assumption may lead to certain differences when compared with a uniform distribution over the entire area, making the calibration slightly different from those reported by 3GPP. 
    \item For the monostatic sensing case, the coefficients generated in steps 2-6 and 11-12 are identical for the STX-SPST and SPST-SRX links, as they correspond to the same physical device. However, special attention is required for steps 7 and 13, where the angle-related and phase-related coefficients do not directly match between the STX-SPST and SPST-SRX links. Indeed, in step 7, the AoDs and AoAs at the STX-SPST link correspond to the AoAs and AoDs at the SPST-SRX links in both the azimuth and elevation domains, respectively. In step 13, the random initial phases associated with the polarization components are also not directly aligned. Specifically, the $\phi^{\theta\phi}$ and $\phi^{\phi\theta}$ elements at the STX-SPST link correspond to $\phi^{\phi\theta}$ and $\phi^{\theta\phi}$ elements at the SPST-SRX link, respectively.
    \item In Step 7 of RP channel generation, rays with ZOA below a predefined threshold are discarded. However, some differences are observed when comparing the simulation results with the calibration results reported by 3GPP, since enabling or disabling ZOA filtering can change the calibration results, suggesting that the impact of the ZOA filtering may require further checking. 
    \item After the ISAC channel coefficients are generated, the calibration metrics are first computed. Subsequently, only a subset of STX-SRX pairs is selected for calibration. Specifically, when conducting the calibration, only $N$ STX–SRX pairs with the smallest coupling loss of the targets are chosen to be included for calibration. The value of $N$ depends on the sensing configuration. For TRP-TRP bistatic sensing, $N=1$, while for TRP–TRP monostatic sensing in the Highway scenario under FR1, $N=2$. For all other cases, $N=4$.
\end{enumerate}

The above implementation details indicate that calibration discrepancies may arise from different interpretations of the 3GPP procedure, rather than coding errors alone. 
These observations suggest that calibration results may be sensitive to implementation choices. Therefore, such assumptions should be explicitly documented when developing and validating a 3GPP-compliant ISAC channel simulator. Note that while the above points seem minor, incorrect handling of them may lead to noticeable errors in the calibration results.

\subsection{Calibration Metrics}

The calibration metrics adopted in this work follow those specified in 3GPP TR 38.901. In particular, as 3GPP only provides reference calibration data for coupling loss, delay spread, and angular spread, these metrics are used to verify whether the implemented simulator reproduces the intended statistical channel behavior agreed by 3GPP. 

\subsubsection{Coupling Loss} The calibration of coupling loss consists of two parts: the large-scale coupling loss and full-scale coupling loss. The large-scale coupling loss is composed of pathloss $PL_{dB}$, large-scale RCS component of SPST $\sigma_{\mathrm{RCS},M}$ and shadow fading $SF_{dB}$, and is generally expressed as:
\begin{equation}
\begin{aligned}
L_{\mathrm{LS}}
=&
PL_{dB}(d_1)
+
PL_{dB}(d_2)
+
10\log\!\left(\frac{c^{2}}{4\pi f^{2}}\right)\\
&-
10\log(\sigma_{\mathrm{RCS},M})
+
SF_{dB,1}
+
SF_{dB,2}\quad\text{[dB]}.
\end{aligned}    
\end{equation}
where $c$ is the speed of light, $f$ is the frequency, $d_1$ and $d_2$ and are the propagation distances of STX-SPST and SPST-SRX links, respectively, and $SF_{dB,1}$ and $SF_{dB,2}$ corresponding to shadow fading of STX-SPST and SPST-SRX links, respectively. The large-scale coupling loss determines the level of large-scale attenuation, while the full-scale formulation additionally captures the small-scale fading effect of each propagation path. Hence, the full-scale coupling loss consists of the large-scale coupling loss and the small-scale channel attenuation of each propagation path, which is expressed as:
\begin{equation}
    L_{\mathrm{Full}}=L_{\mathrm{LS}}+20\log_{10}(L_{\mathrm{SM}})\quad\text{[dB]},
\end{equation}
where $L_{\mathrm{SM}}=\sqrt{P_{n}} \mathbf{F}_{rx}^{T} \mathbf{Q}_\mathbf{CPM} \mathbf{F}_{tx}$, where $P_n$ is the per-path power of path $n$, $\mathbf{F}_{tx}$ and $\mathbf{F}_{rx}$ are the transmit and receive antenna field pattern vectors, respectively, and $\mathbf{Q}_\mathbf{CPM}$ is the combined CPM of the transmitter, sensing target, and receiver.

\subsubsection{Delay Spread and Angular Spread}

The definitions of the delay spread (DS) and angular spread (AS) follow those provided in 3GPP TR 38.901 and are identical to those used in communication channel calibration. These metrics are computed from the generated channel coefficients and characterize the temporal and angular dispersion of the channel, respectively. They are commonly used to validate the consistency of the implemented ISAC channel model.

\begin{figure}[t]
\centering
\vspace{0.1in}
\begin{subfigure}[t]{0.45\linewidth}
\centering
\includegraphics[width=\linewidth]{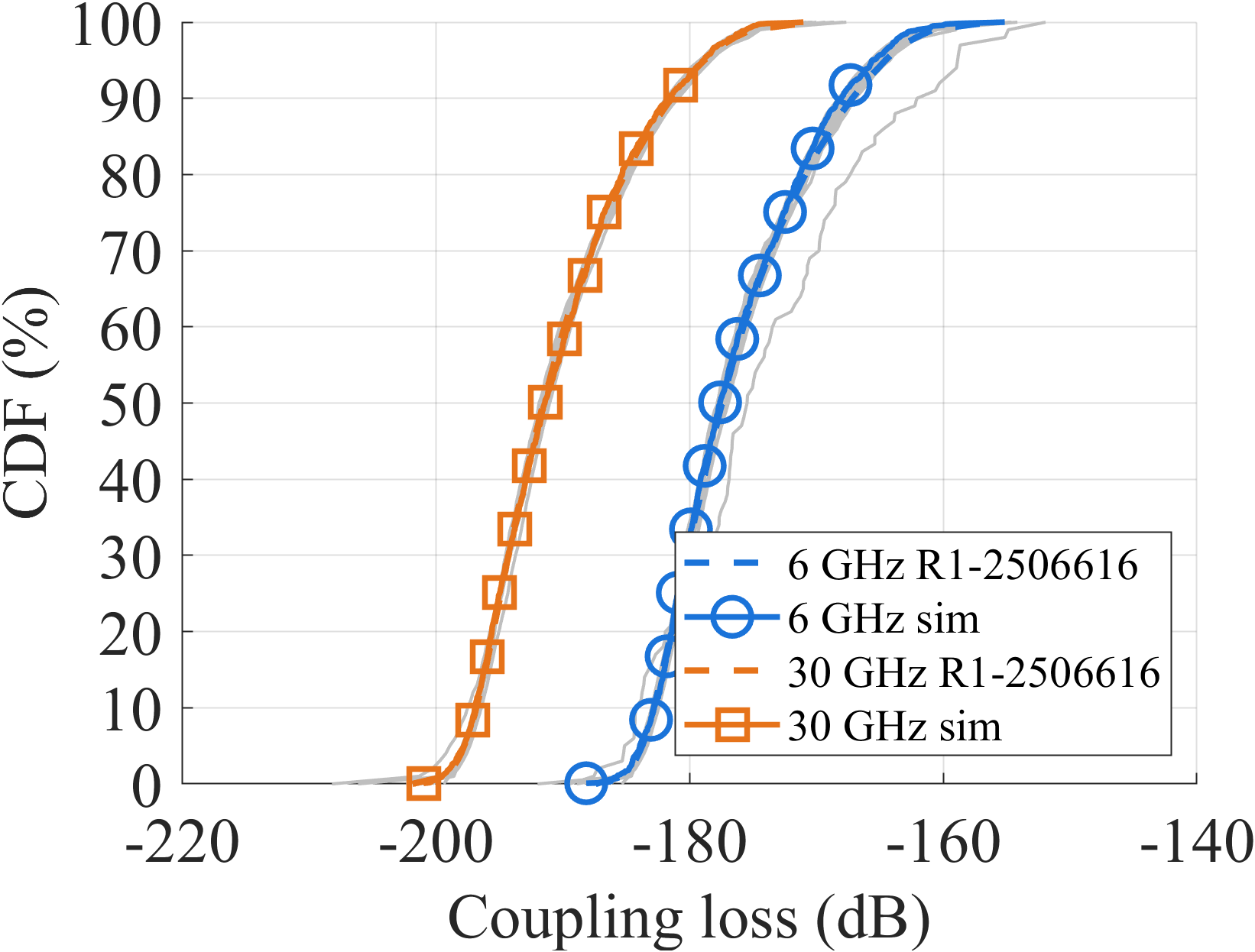}
\caption{Large-scale coupling loss.}
\label{fig:option 1 LS_TRP_mono_t_CouplingLoss}
\end{subfigure}
\hfill
\begin{subfigure}[t]{0.45\linewidth}
\includegraphics[width=\linewidth]{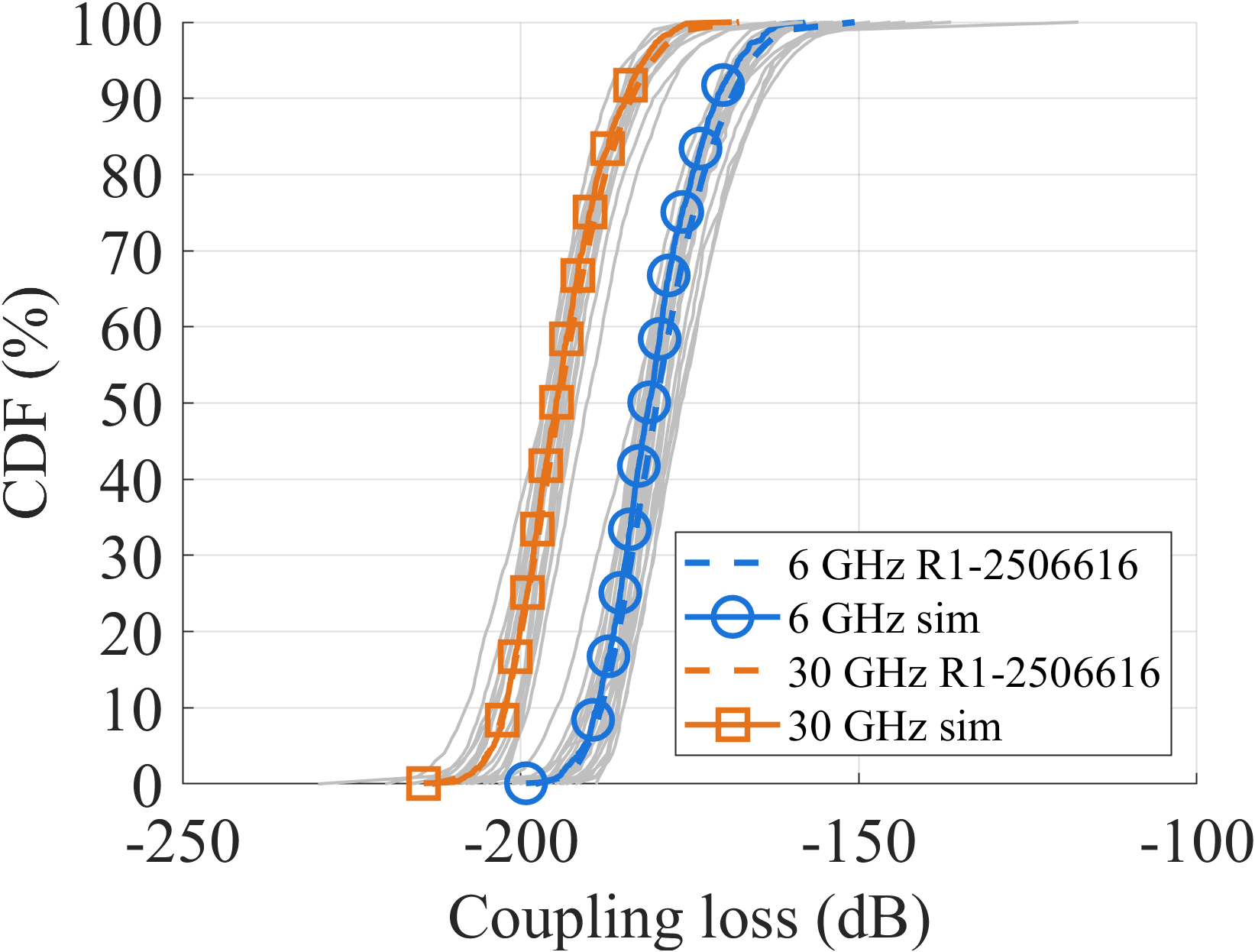}
\caption{Full-scale coupling loss.}
\label{fig:option 1 Full_TRP_mono_t_CouplingLoss}
\end{subfigure}

\begin{subfigure}[t]{0.45\linewidth}
\includegraphics[width=\linewidth]{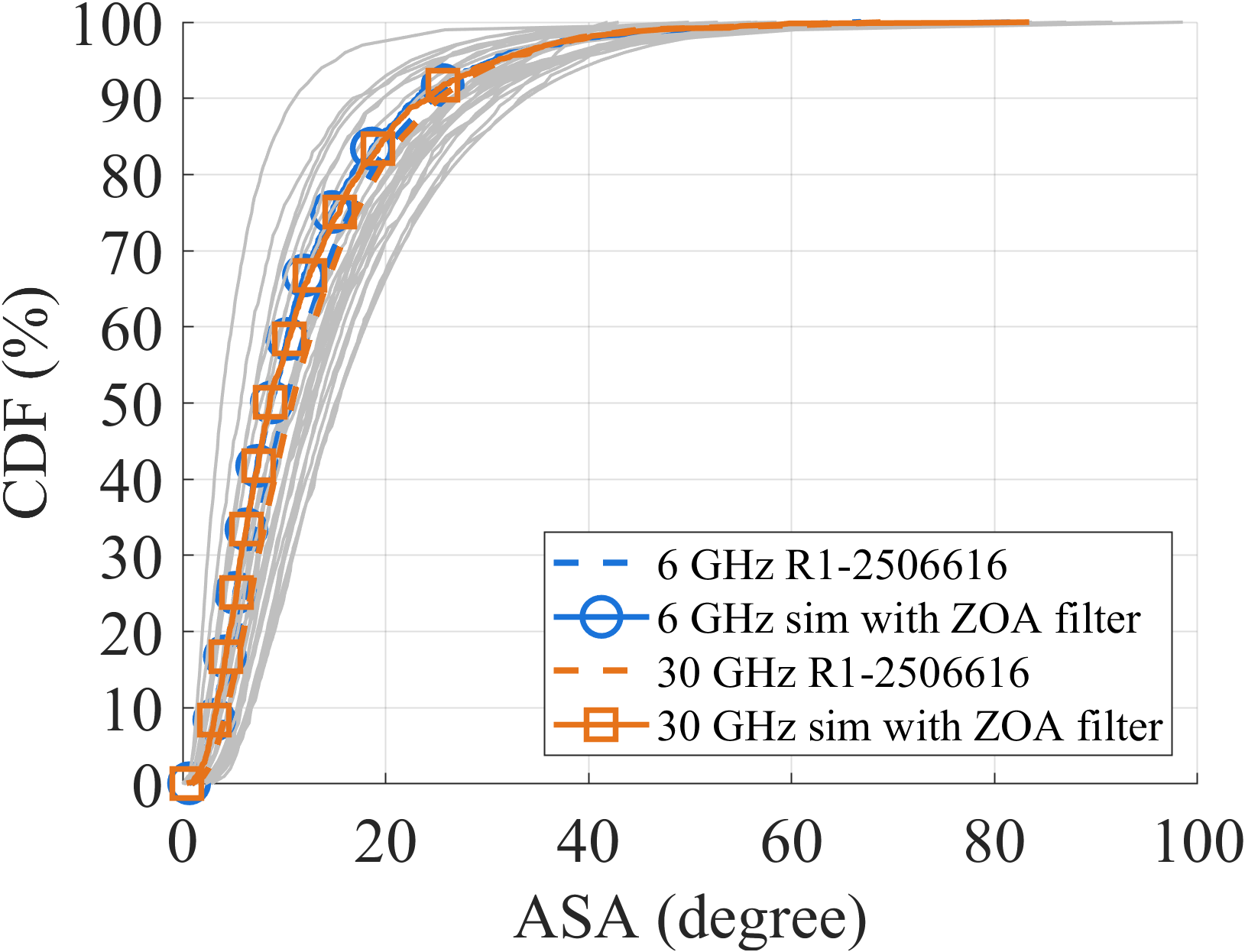}
\caption{ASA.}
\label{fig:option 1 Full_TRP_mono_t_ASA}
\end{subfigure}
\hfill
\begin{subfigure}[t]{0.45\linewidth}
\includegraphics[width=\linewidth]{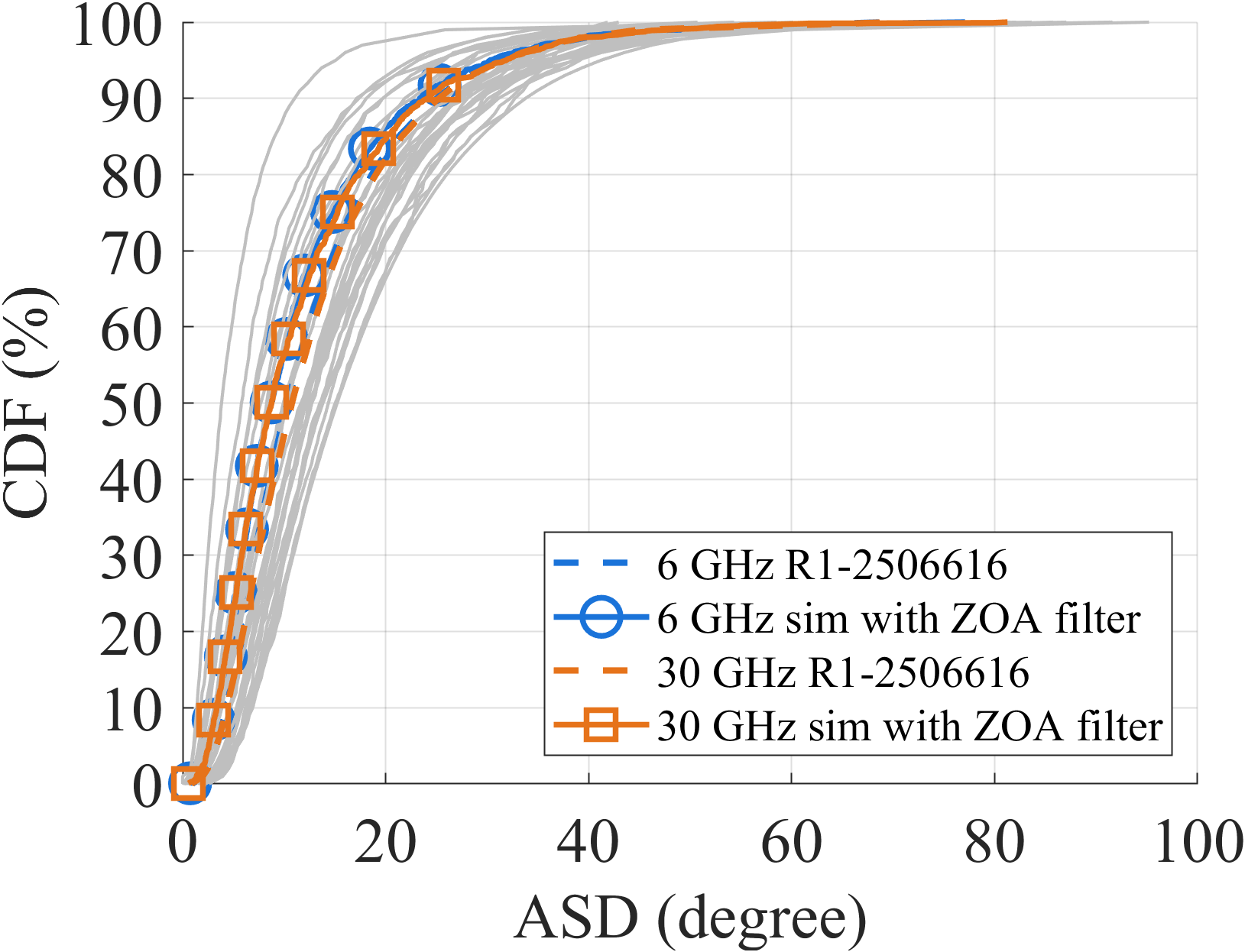}
\caption{ASD.}
\label{fig:option 1 Full_TRP_mono_t_ASD}
\end{subfigure}

\begin{subfigure}[t]{0.45\linewidth}
\includegraphics[width=\linewidth]{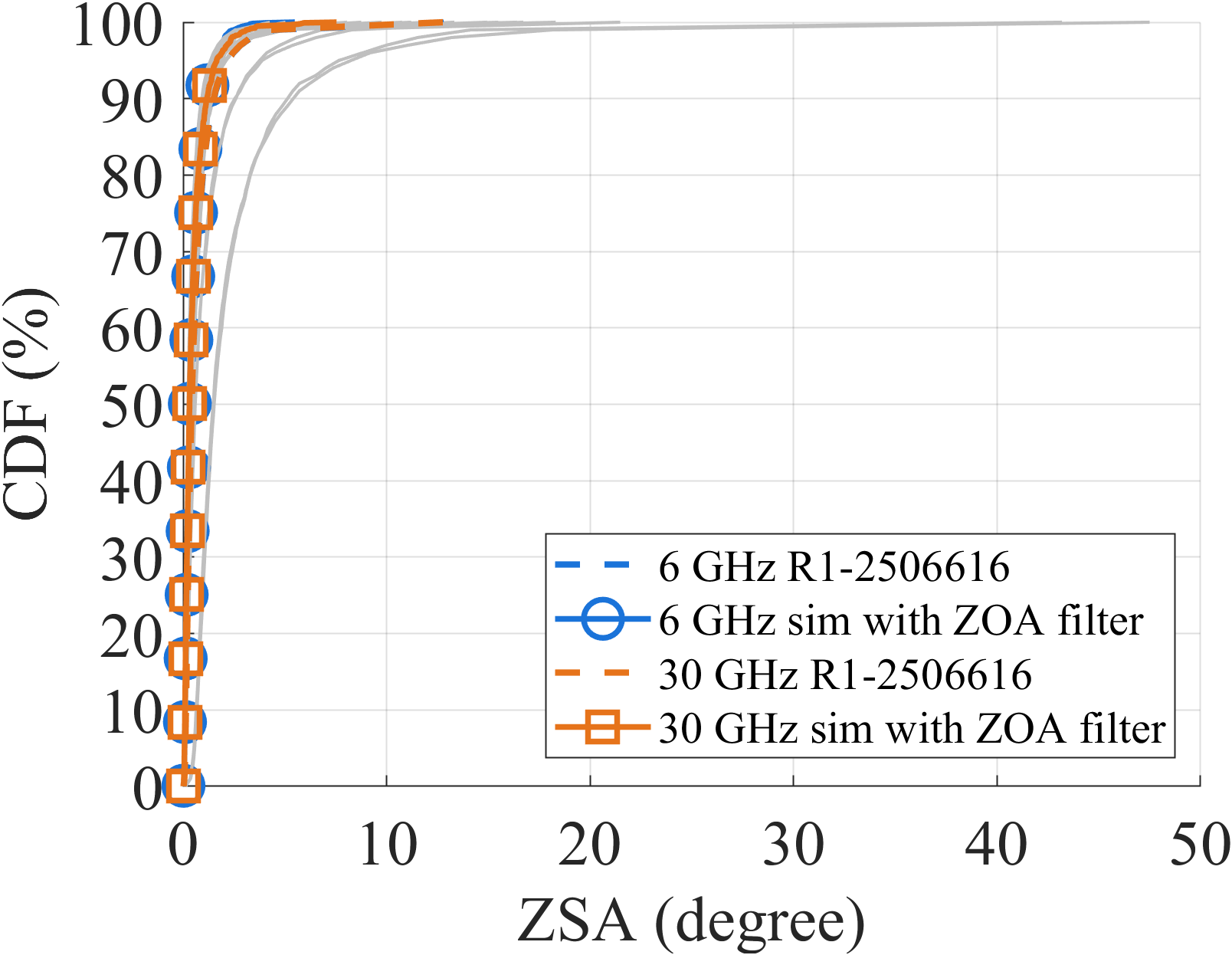}
\caption{ZSA.}
\label{fig:option 1 Full_TRP_mono_t_ZSA}
\end{subfigure}
\hfill
\begin{subfigure}[t]{0.45\linewidth}
\includegraphics[width=\linewidth]{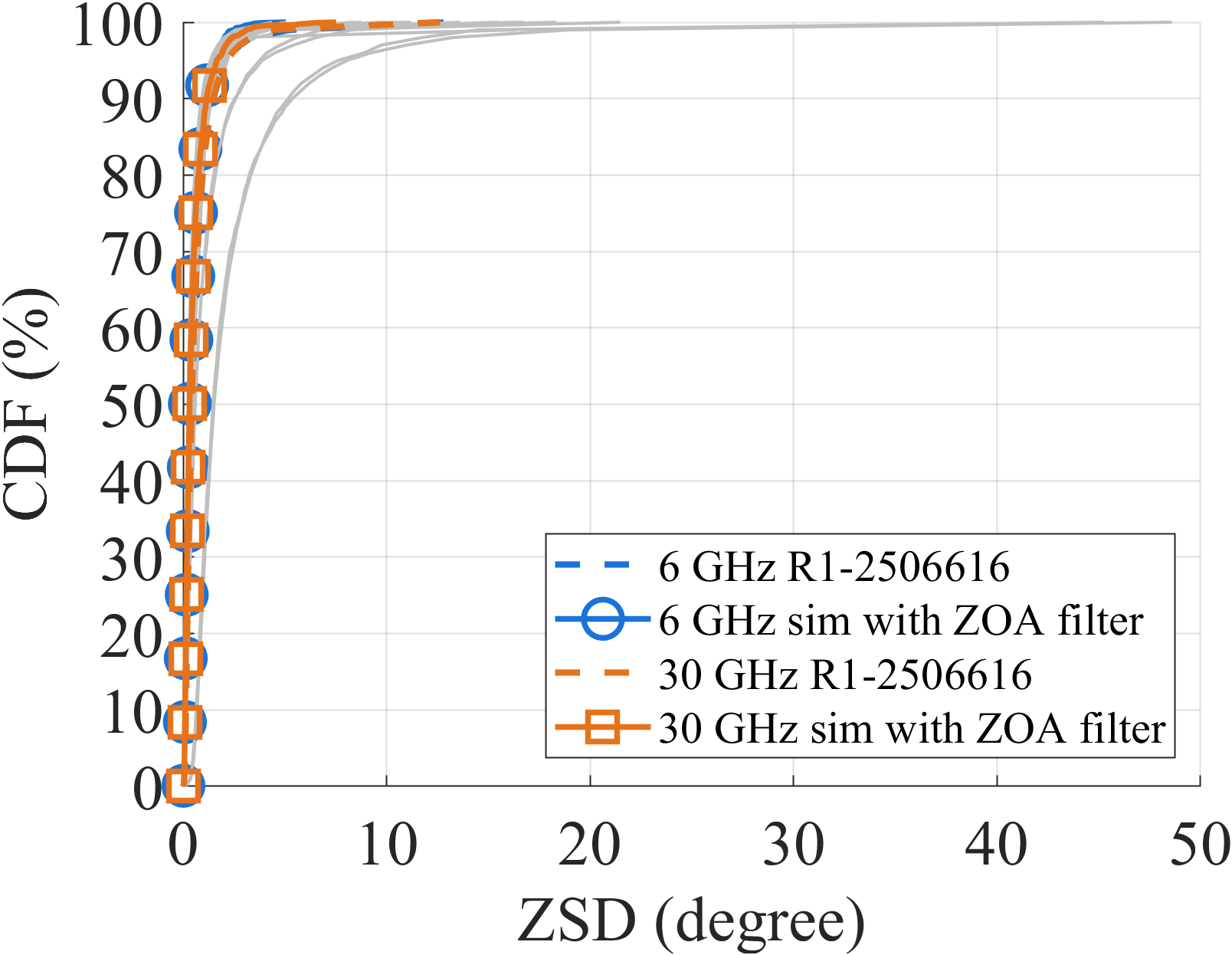}
\caption{ZSD.}
\label{fig:option 1 Full_TRP_mono_t_ZSD}
\end{subfigure}

\begin{subfigure}[t]{0.45\linewidth}
\includegraphics[width=\linewidth]{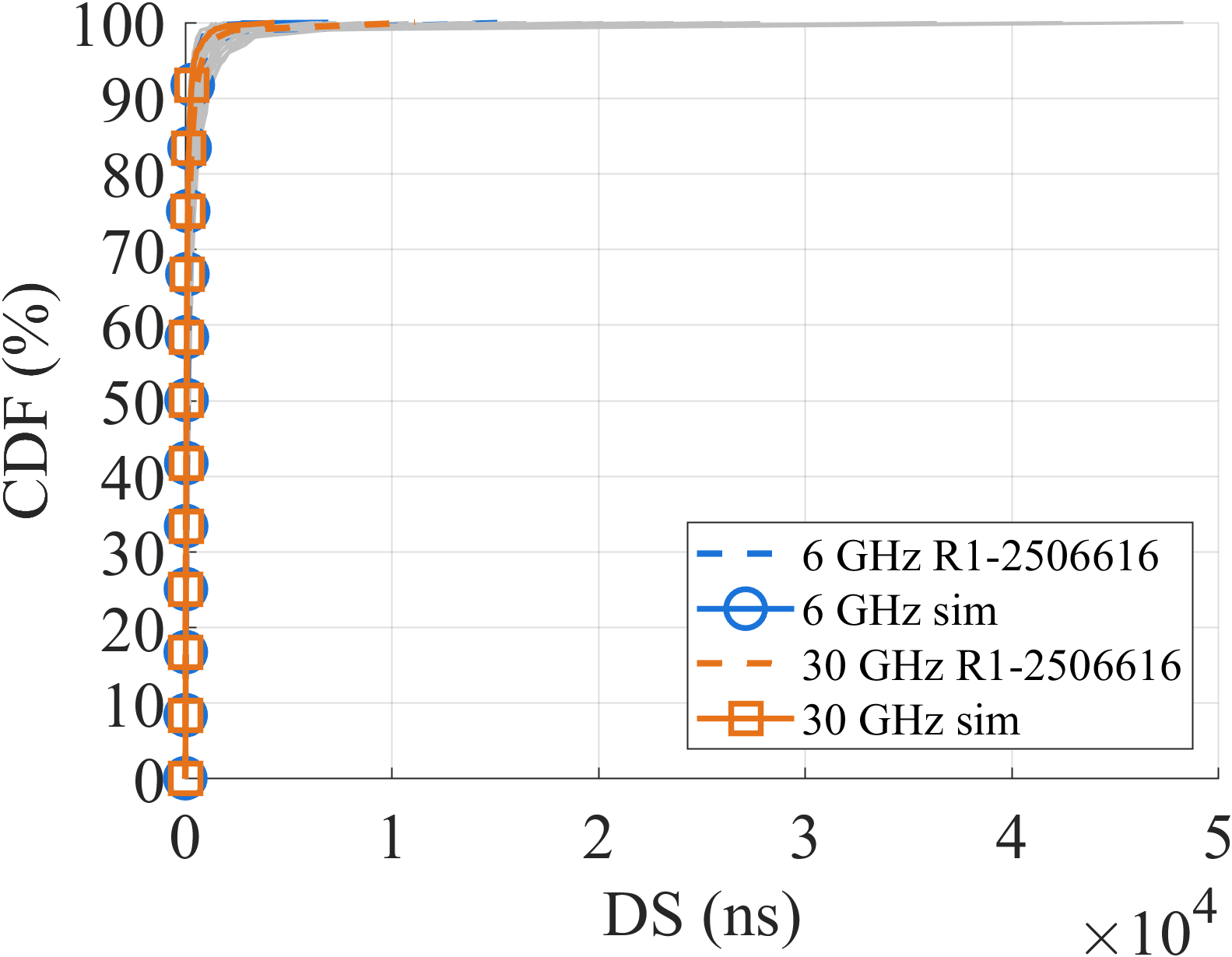}
\caption{DS.}
\label{fig:option 1 Full_TRP_mono_t_DS}
\end{subfigure}

\caption{Target channel calibrations for UMa-AV scenario with parameters in \cite[Table 7.9.6.1-1, 7.9.6.2-1]{3gpp38901v1910} and option 1.}
\label{fig:option 1 target channel}
\vspace{-10pt}
\end{figure}

\begin{figure}[t]
\centering
\vspace{0.1in}
\begin{subfigure}[t]{0.45\linewidth}
\centering
\includegraphics[width=\linewidth]{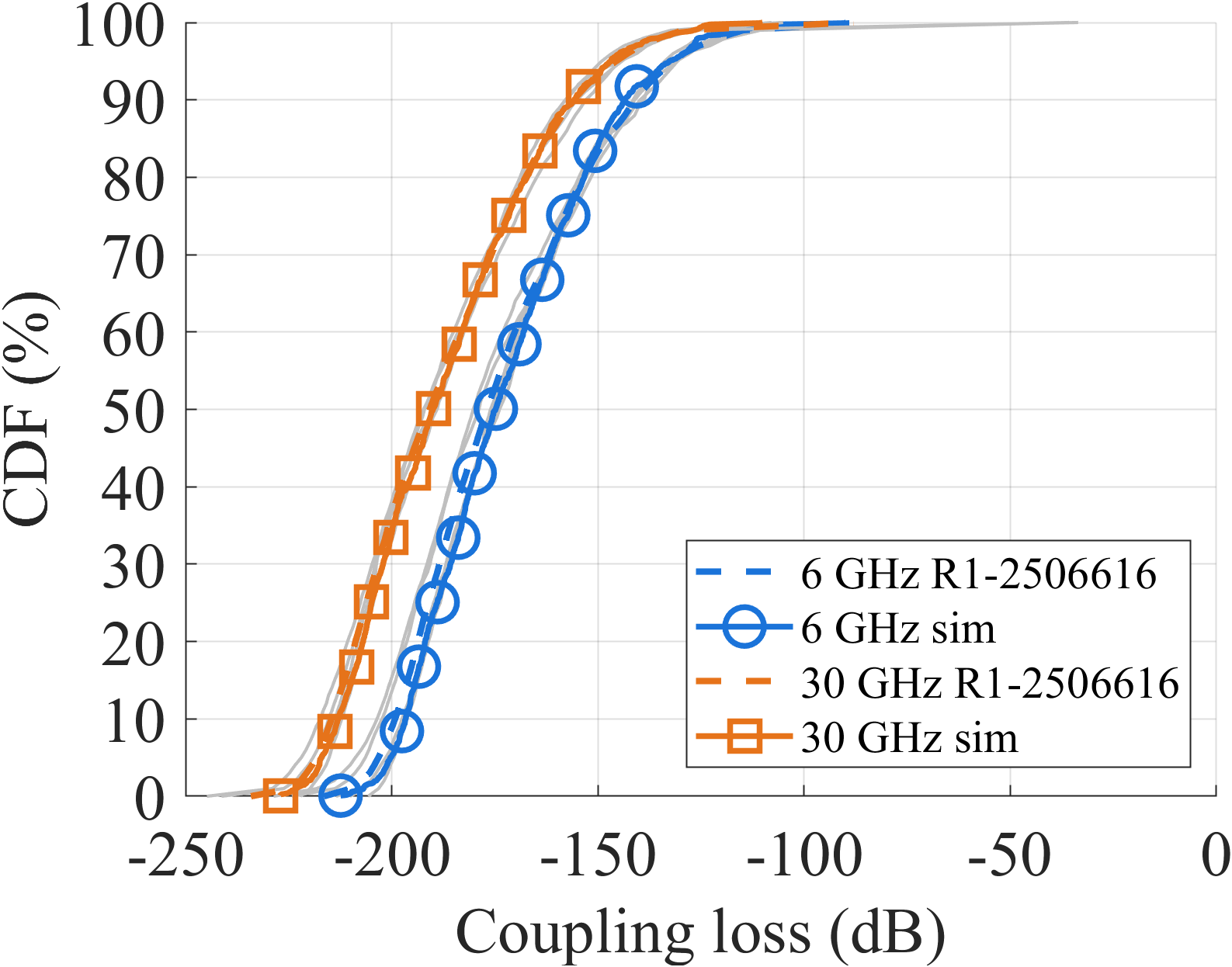}
\caption{Large-scale coupling loss.}
\label{fig:comm_2_sen_2 LS_TRP_mono_t_CouplingLoss}
\end{subfigure}
\hfill
\begin{subfigure}[t]{0.45\linewidth}
\includegraphics[width=\linewidth]{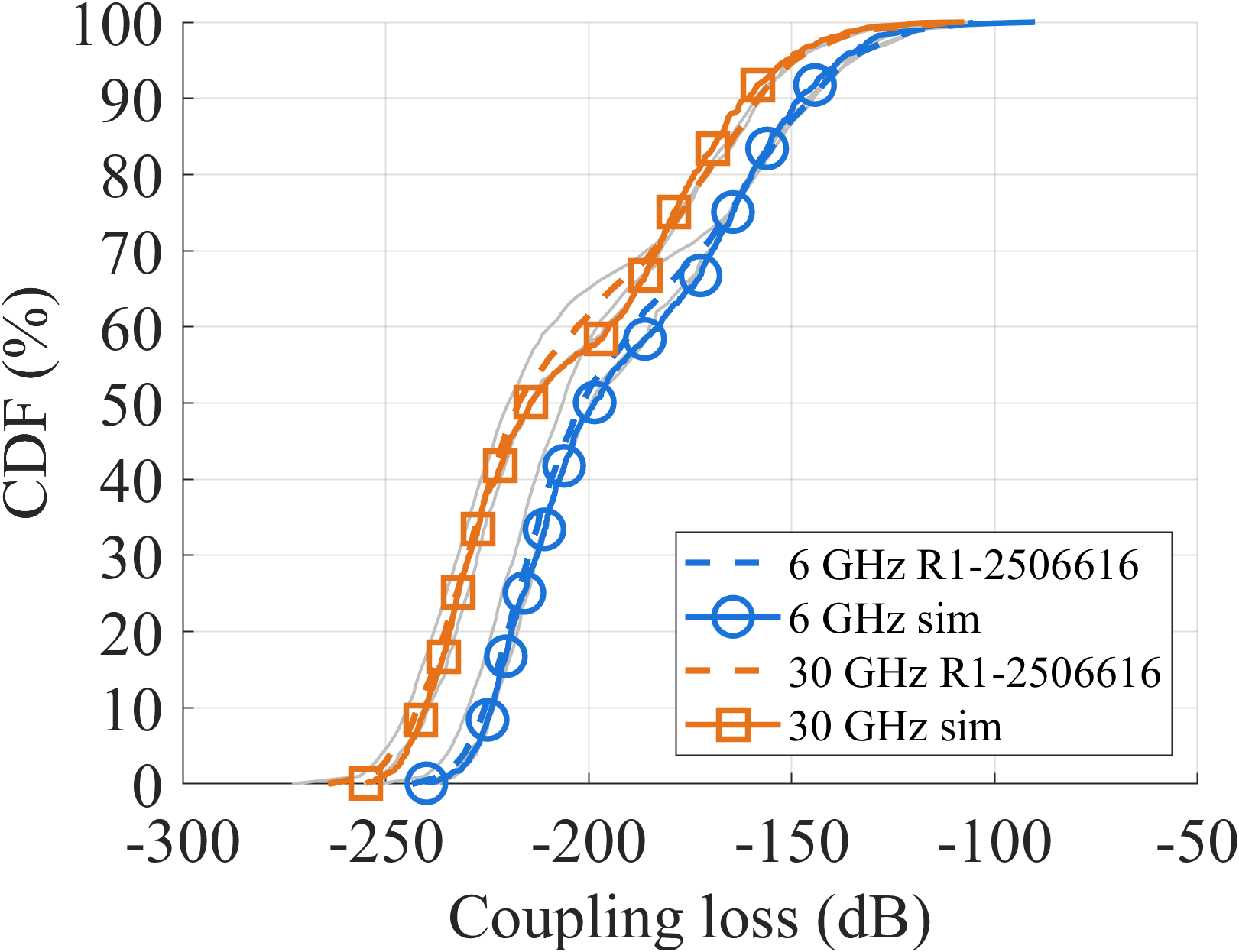}
\caption{Full scale coupling loss.}
\label{fig:comm_2_sen_2 Full_TRP_mono_t_CouplingLoss}
\end{subfigure}

\begin{subfigure}[t]{0.45\linewidth}
\includegraphics[width=\linewidth]{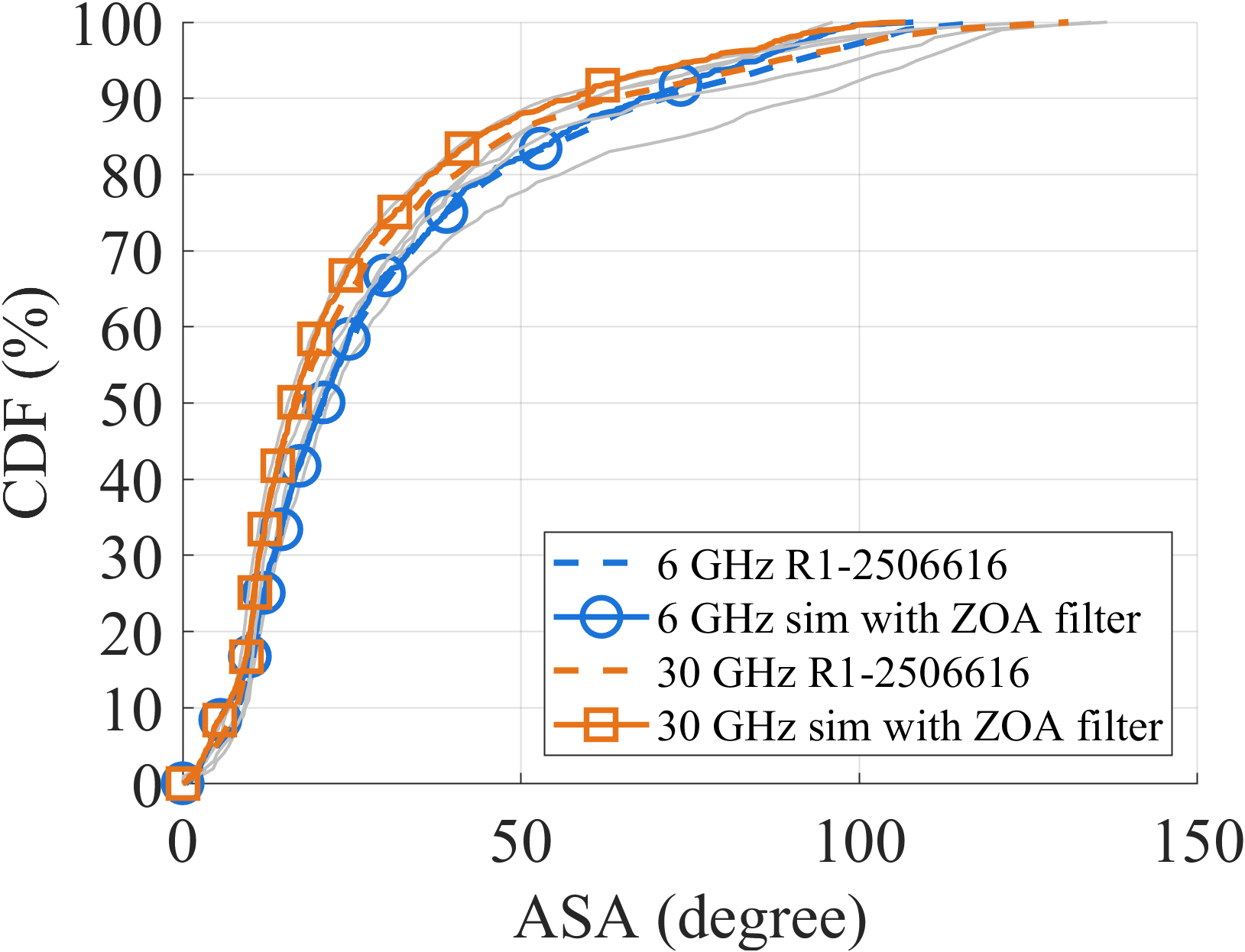}
\caption{ASA.}
\label{fig:comm_2_sen_2 Full_TRP_mono_t_ASA}
\end{subfigure}
\hfill
\begin{subfigure}[t]{0.45\linewidth}
\includegraphics[width=\linewidth]{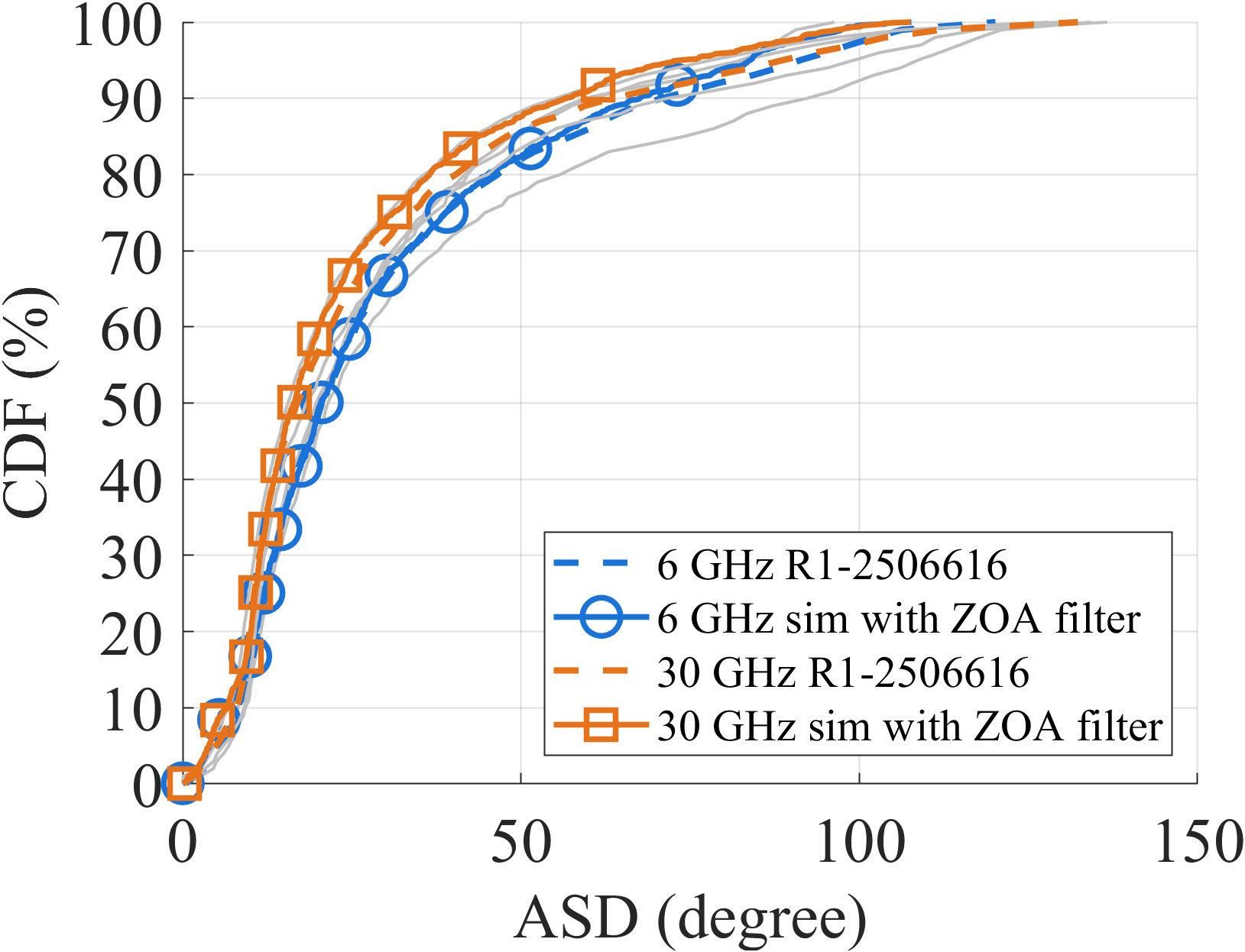}
\caption{ASD.}
\label{fig:comm_2_sen_2 Full_TRP_mono_t_ASD}
\end{subfigure}

\begin{subfigure}[t]{0.45\linewidth}
\includegraphics[width=\linewidth]{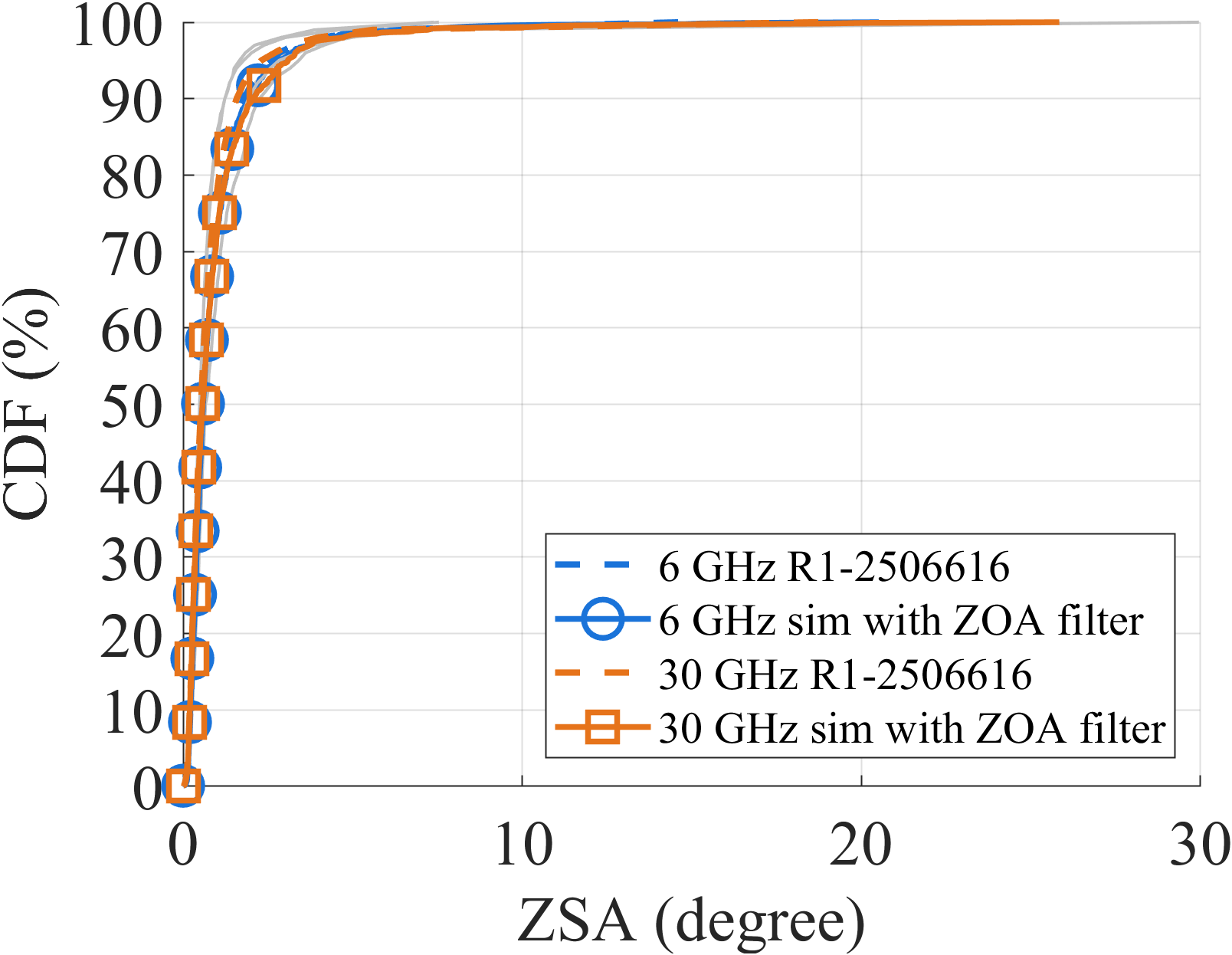}
\caption{ZSA.}
\label{fig:comm_2_sen_2 Full_TRP_mono_t_ZSA}
\end{subfigure}
\hfill
\begin{subfigure}[t]{0.45\linewidth}
\includegraphics[width=\linewidth]{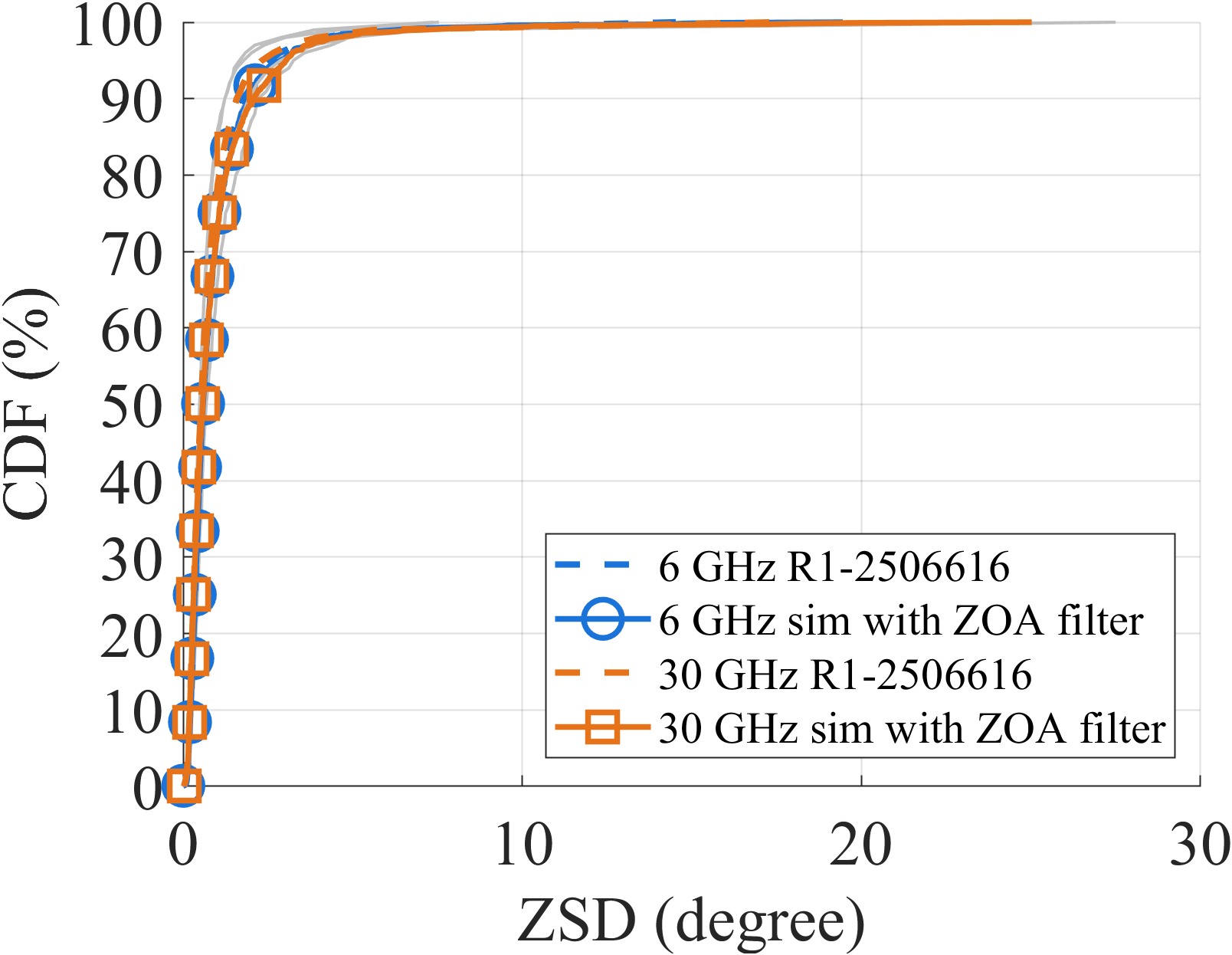}
\caption{ZSD.}
\label{fig:comm_2_sen_2 Full_TRP_mono_t_ZSD}
\end{subfigure}

\begin{subfigure}[t]{0.45\linewidth}
\includegraphics[width=\linewidth]{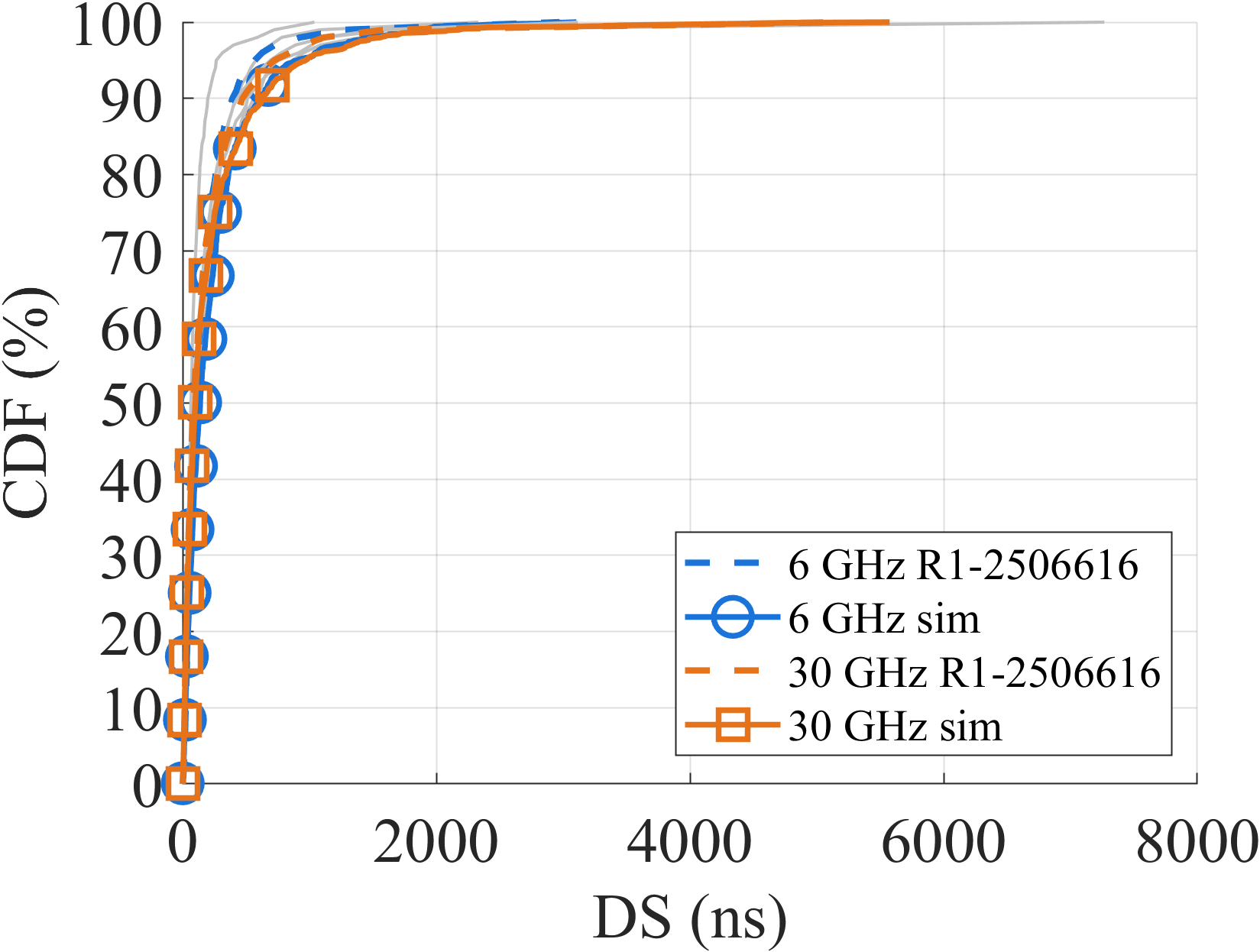}
\caption{DS.}
\label{fig:comm_2_sen_2 Full_TRP_mono_t_DS}
\end{subfigure}

\caption{Target channel calibrations for UMi Human scenario with the parameters in \cite[Table 7.9.6.1-2, 7.9.6.2-2]{3gpp38901v1910}.}
\label{fig:comm_2_sen_2 target channel}
\vspace{-10pt}
\end{figure}

\section{Calibration Results}

In this section, we provide the calibration results of the ISAC channel model simulator of 3GPP TR 38.901. The results are compared with the reference results reported in 3GPP R1-2506616~\cite{R1-2506616}. The calibration in this work focuses on the monostatic configuration. The procedure can be extended to bistatic configurations. Note that unless otherwise indicated, option 2 is adopted in step 9 of the target channel generation. Due to space limitations, only representative results are presented and discussed here. The complete set of calibration results for other sensing types and scenarios is available on GitHub.


\subsection{Target Channel Calibrations}

\begin{figure}[t]
\centering
\vspace{0.1in}
\begin{subfigure}[t]{0.45\linewidth}
\centering
\includegraphics[width=\linewidth]{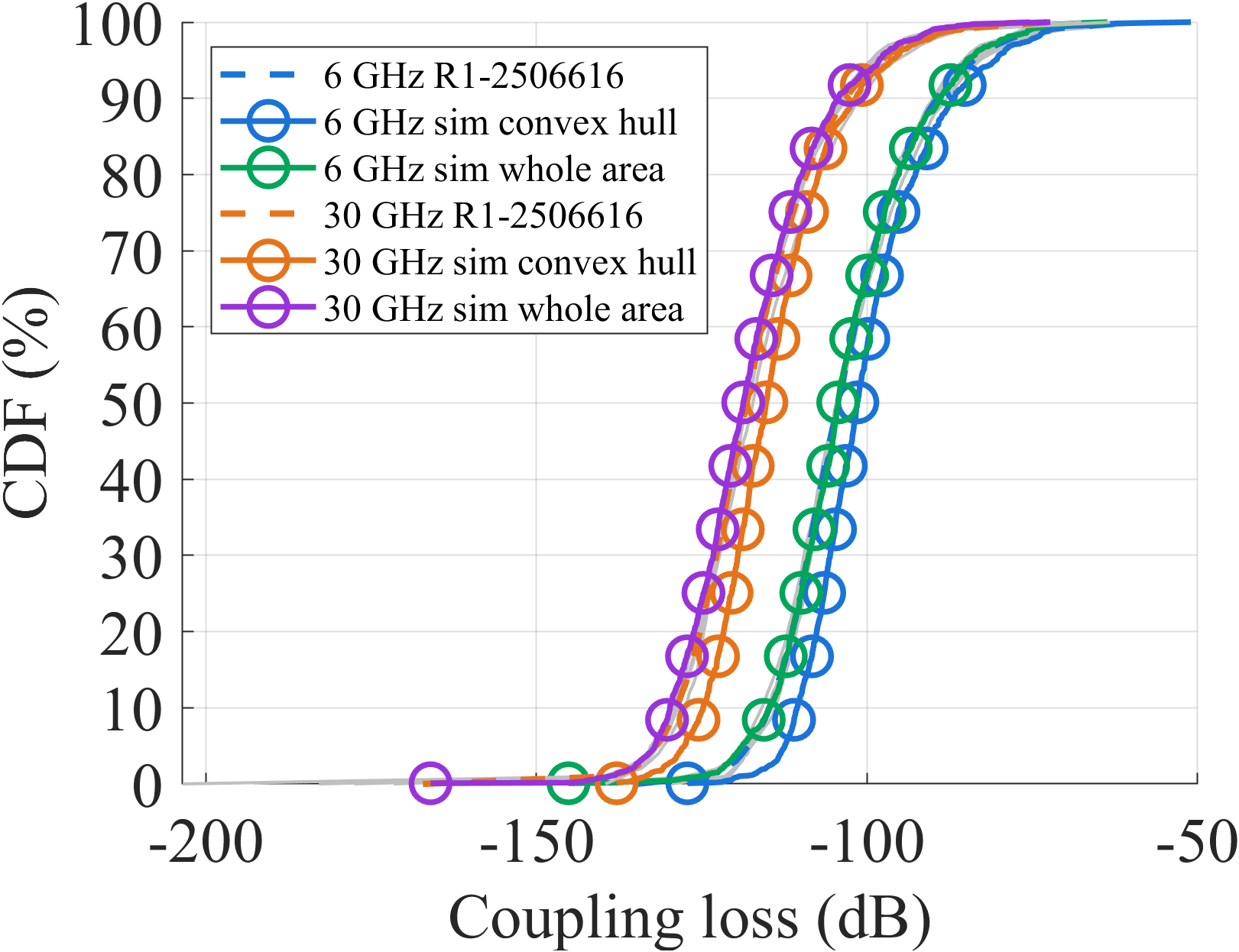}
\caption{Large-scale coupling loss.}
\label{fig:comm_3_sen_2 LS_TRP_mono_t_CouplingLoss}
\end{subfigure}
\hfill
\begin{subfigure}[t]{0.45\linewidth}
\includegraphics[width=\linewidth]{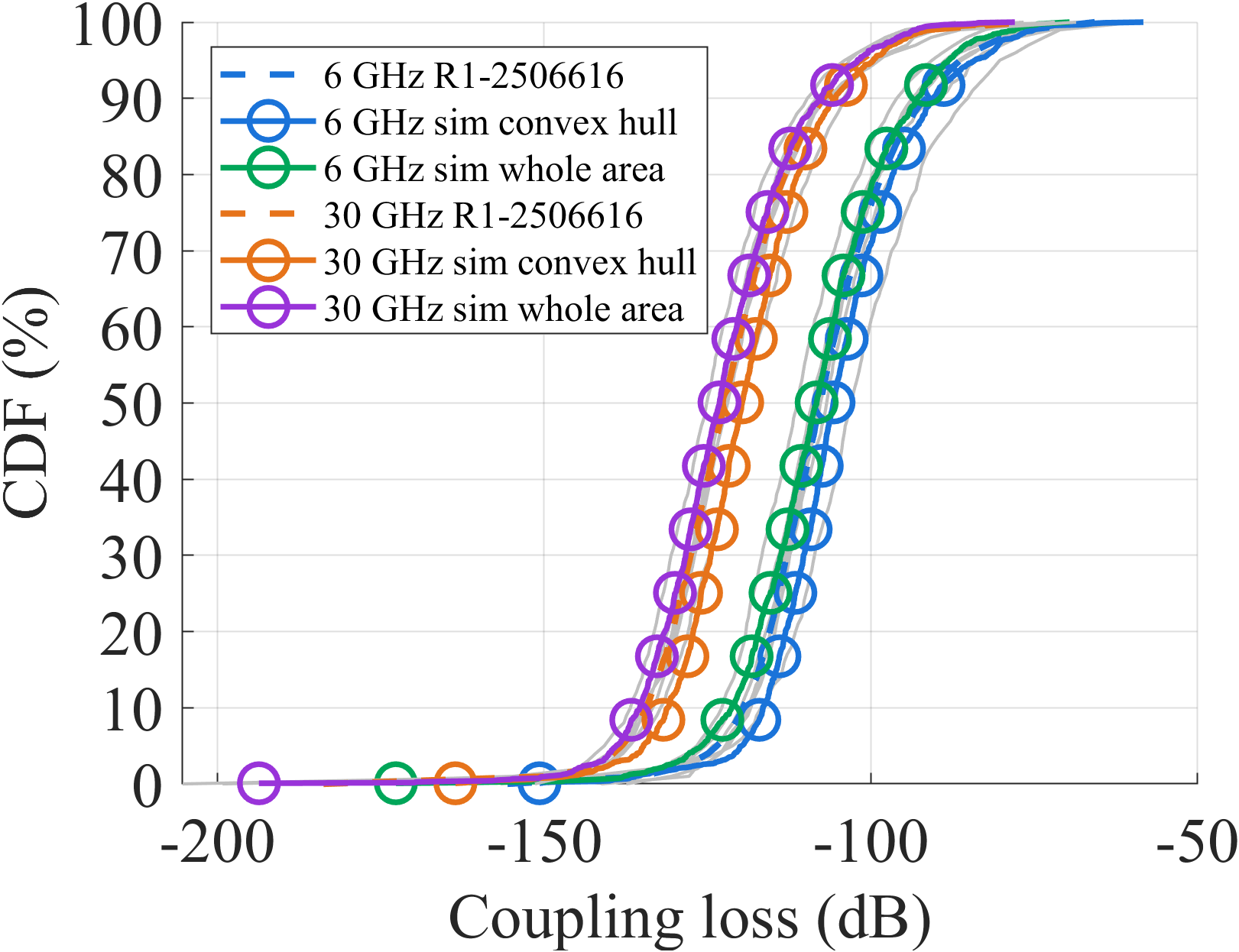}
\caption{{Full-scale coupling loss.}}
\label{fig:comm_3_sen_2 Full_TRP_mono_t_CouplingLoss}
\end{subfigure}




\caption{Target channel calibrations for InH Human scenario with parameters in \cite[Table 7.9.6.1-2, 7.9.6.2-2]{3gpp38901v1910}.}
\label{fig:comm_3_sen_2 target channel}
\end{figure}

\begin{figure}[t]
\centering
\begin{subfigure}[t]{0.45\linewidth}
\centering
\includegraphics[width=\linewidth]{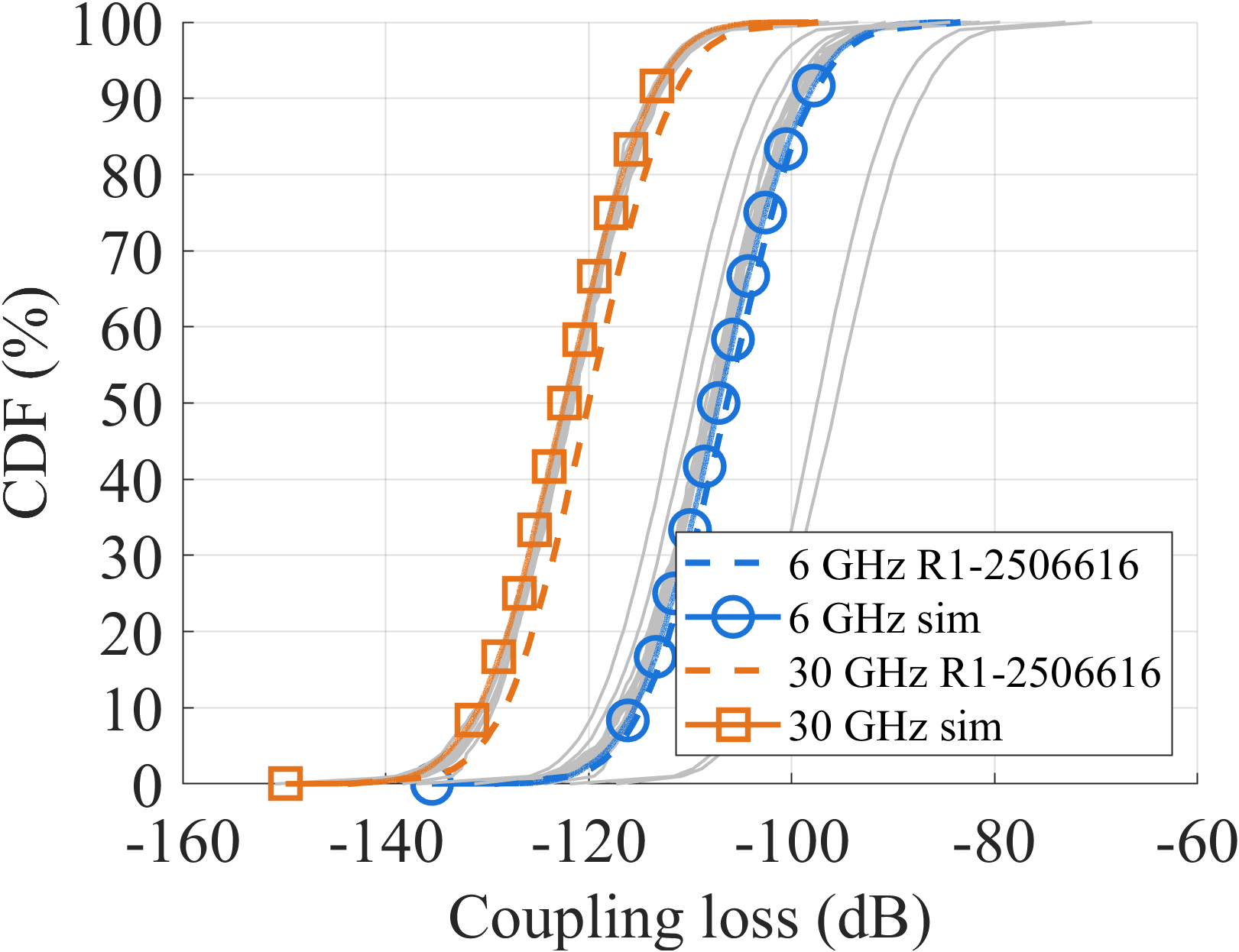}
\caption{Large-scale coupling loss.}
\label{fig:option 1 LS_TRP_mono_b_CouplingLoss}
\end{subfigure}
\hfill
\begin{subfigure}[t]{0.45\linewidth}
\includegraphics[width=\linewidth]{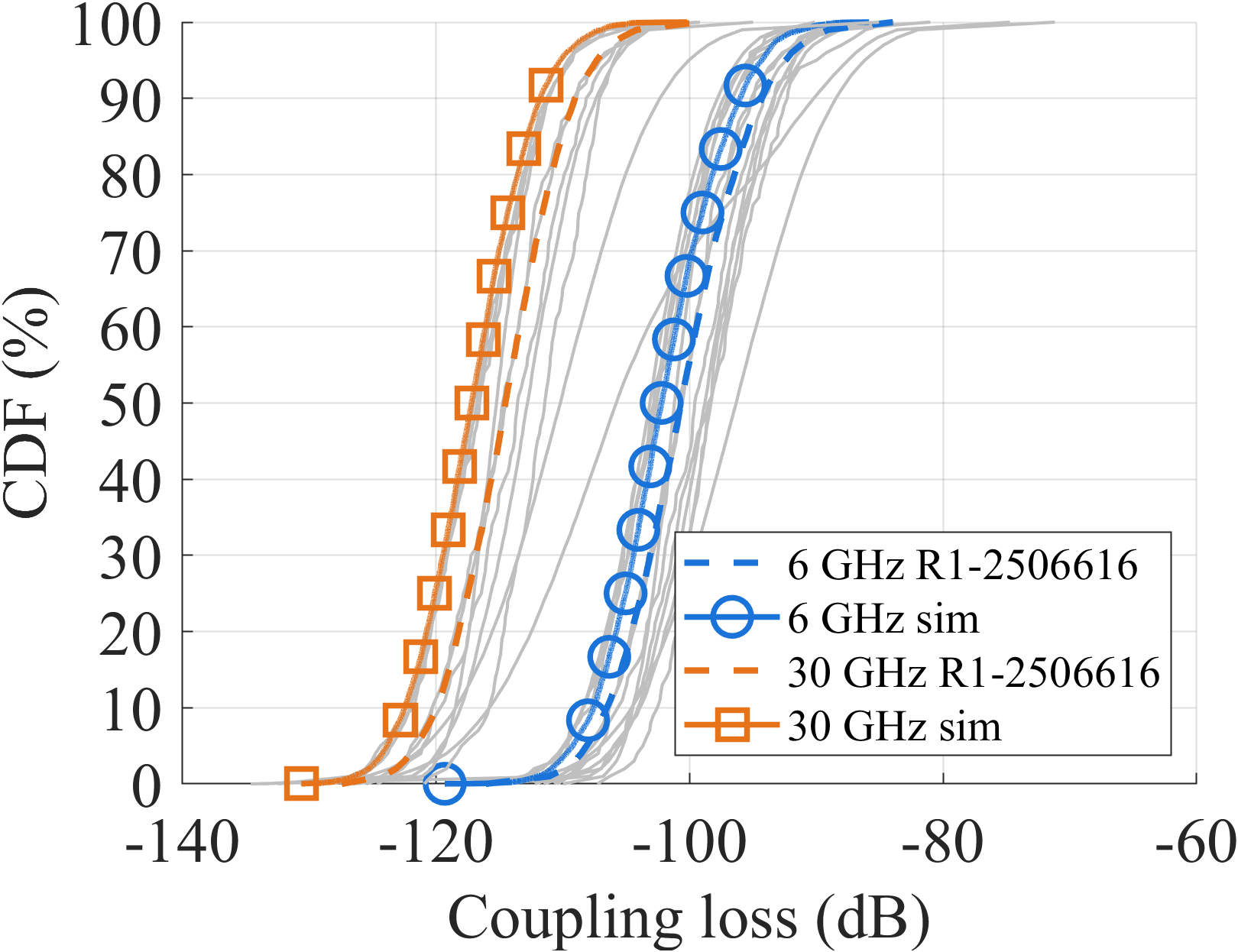}
\caption{Full-scale coupling loss.}
\label{fig:option 1 Full_TRP_mono_b_CouplingLoss}
\end{subfigure}

\begin{subfigure}[t]{0.45\linewidth}
\includegraphics[width=\linewidth]{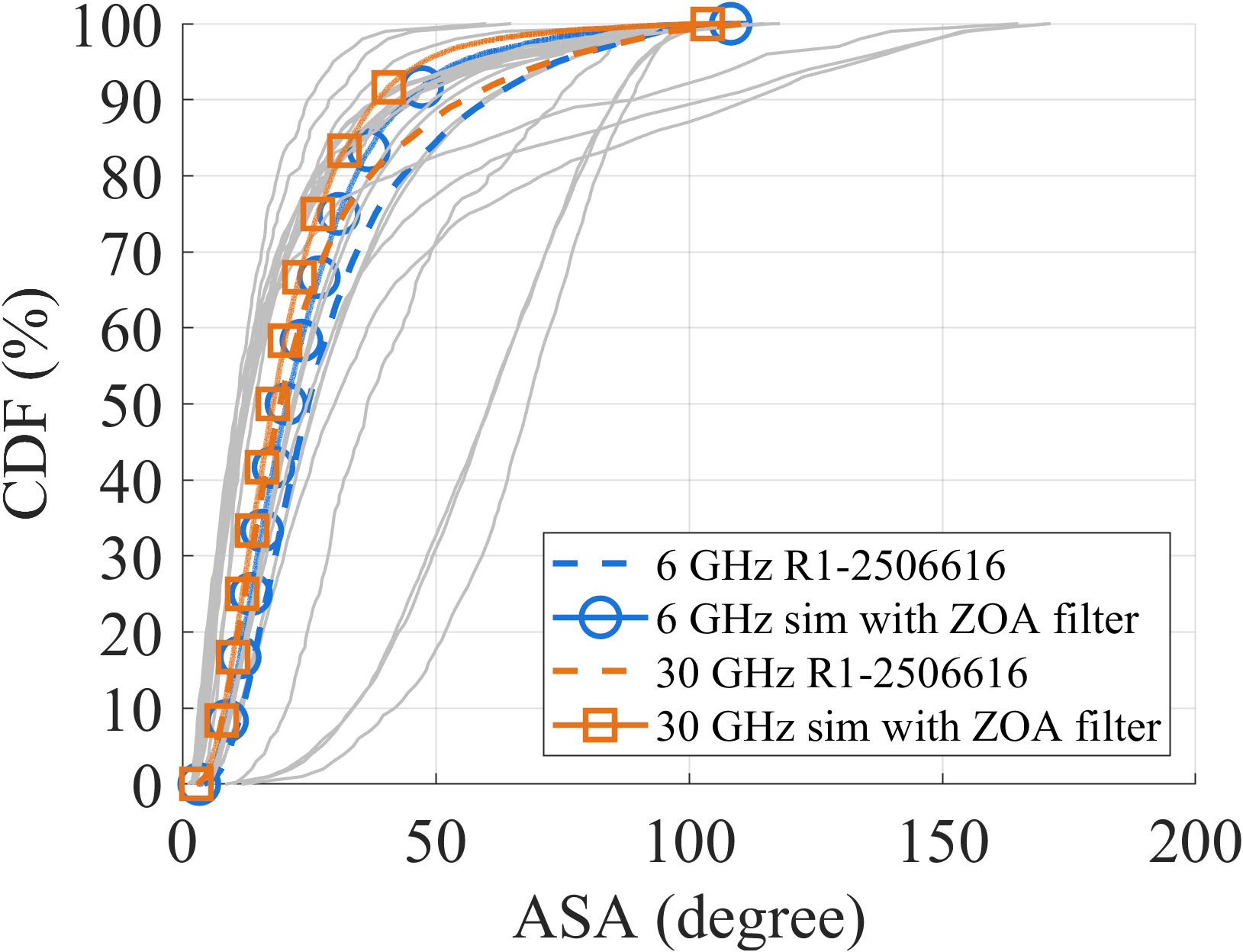}
\caption{ASA.}
\label{fig:option 1 Full_TRP_mono_b_ASA}
\end{subfigure}
\hfill
\begin{subfigure}[t]{0.45\linewidth}
\includegraphics[width=\linewidth]{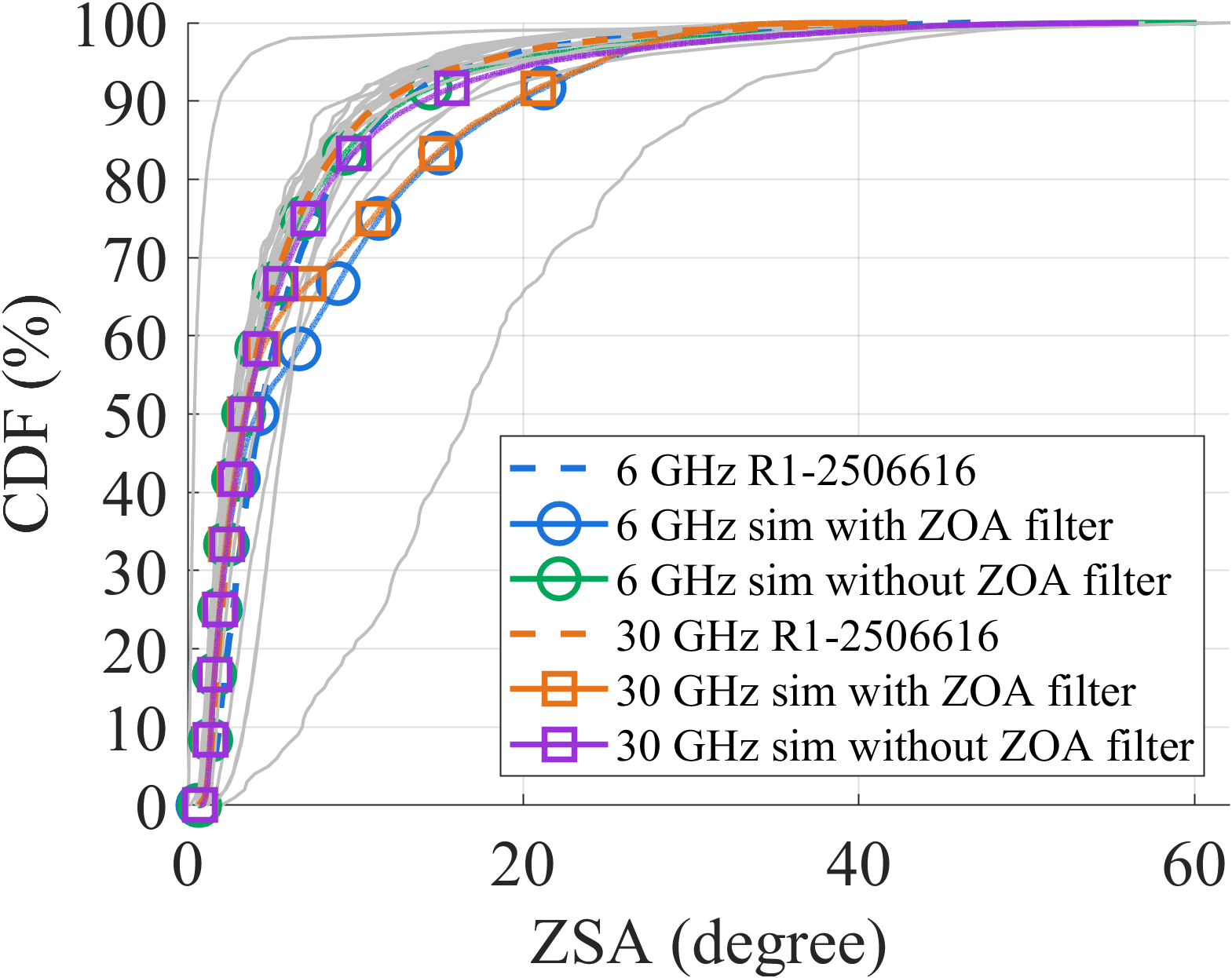}
\caption{ZSA.}
\label{fig:option 1 Full_TRP_mono_b_ZSA}
\end{subfigure}

\begin{subfigure}[t]{0.45\linewidth}
\includegraphics[width=\linewidth]{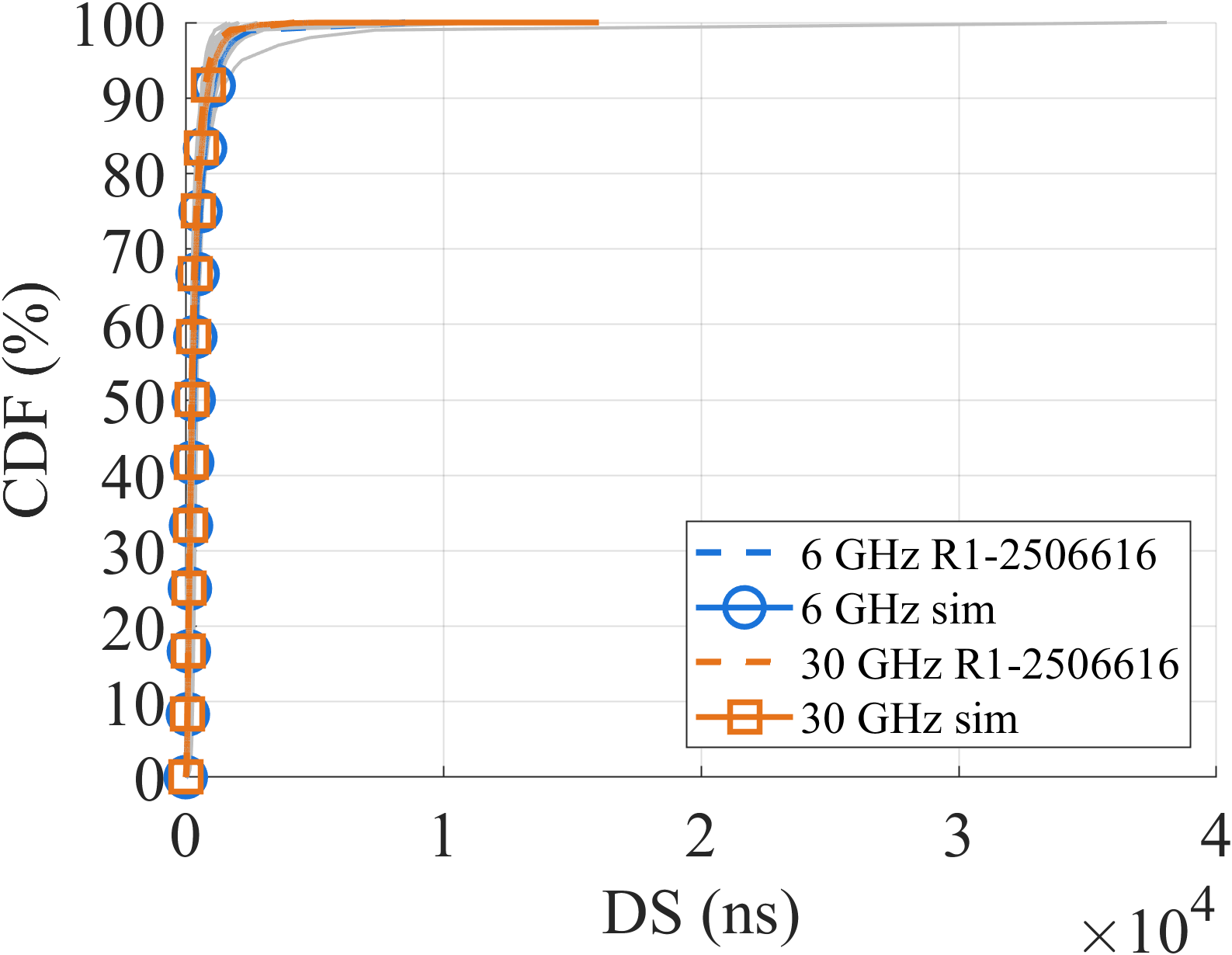}
\caption{DS.}
\label{fig:option 1 Full_TRP_mono_b_DS}
\end{subfigure}

\caption{Background channel calibrations for UMa-AV with parameters in \cite[Table 7.9.6.1-1, 7.9.6.2-1]{3gpp38901v1910} and option 1.}
\vspace{-15pt}
\label{fig:option 1 background channel}
\end{figure}

The target channel calibration results for UMa-AV and UMi Human scenarios are provided in Figs. \ref{fig:option 1 target channel} and \ref{fig:comm_2_sen_2 target channel}, where the gray curves are results from different companies and the dashed curves are the average of results of all companies. It can be observed that the calibration results are generally validated by comparing the CDFs of several key channel parameters. In addition, the results show that our developed simulator closely matches those reported by 3GPP, indicating successful calibration. However, it can also be observed that some figures contain diverse gray curves. This is because 3GPP only provides the results reported by companies, and there is no guarantee that all reported curves are perfectly consistent. Hence, in some cases, the results provided by different companies may differ.


Fig.~\ref{fig:comm_3_sen_2 target channel} presents the coupling loss calibration results of InH-Human scenario as a representative indoor case. It can be observed that the simulation results show good agreement with the reference results of 3GPP. However, it can also be observed in Fig.~\ref{fig:comm_3_sen_2 LS_TRP_mono_t_CouplingLoss} and Fig.~\ref{fig:comm_3_sen_2 Full_TRP_mono_t_CouplingLoss} that when different target distributions are used, i.e., a uniform distribution within the convex hull of the TRP deployment or a uniform distribution over the entire area, the coupling loss results are slightly different. While it is found that the latter setting provides a better match to the reference calibration results in 3GPP, according to~\cite[Table~7.9.6.1-2]{3gpp38901v1910}, the calibration setting should follow the former assumption. Such inconsistencies can occur when comparing with results reported in 3GPP and therefore require special attention.

\subsection{Background Channel Calibrations}

Fig.~\ref{fig:option 1 background channel} shows the calibration results for the background channels in the UMa-AV scenario, where the AoD results are omitted for brevity because AoA and AoD have identical statistics in the background channels. It can again be observed that our calibrated results show good agreement with those provided by 3GPP companies. Nevertheless, for Fig.~\ref{fig:option 1 Full_TRP_mono_b_ZSA}, a noticeable difference in the ZSA results can be observed depending on whether ZOA-based ray filtering is adopted, and the results without ZOA filtering appear to provide a better match to the reference calibration results. However, according to~\cite[Table~7.9.4.2, Step~4]{3gpp38901v1910}, ZOA-based ray filtering for clusters with ZOA less than $50$, $80$, and $90$ degrees should be removed for the UMi, UMa, and RMa scenarios, respectively, making this result somewhat inconsistent with the description in \cite{3gpp38901v1910}. Interestingly, such a difference is not observed in some other scenarios, e.g., the UMi scenario. This might be because the UMi scenario has a smaller angle spread. Hence, the impact of the ZOA filtering is generally minor. Finally, compared with the results of the target channel, the results for the background channel show larger variations across different companies and organizations. Therefore, some gray curves exhibit significant discrepancies. Nevertheless, our simulated results still remain consistent with some of the individual reference results. 

\section{Conclusions}


In this paper, we presented the implementation of a 3GPP-compliant ISAC channel simulator based on TR 38.901 and described its calibration procedure. By comparing the simulated CDFs with the reference results reported by companies in 3GPP, we verified that the developed simulator reproduces the intended statistical behavior of the 3GPP ISAC channel model. We also discussed several implementation-dependent details that may affect calibration outcomes, highlighting the importance of clearly documenting implementation assumptions for reproducible and interoperable ISAC channel simulations. To facilitate reproducibility and further research, the developed simulator, together with the relevant datasets and calibration results, has been released as an open-source project on GitHub.

\bibliographystyle{IEEEtran}
\bibliography{references}

\end{document}